\documentclass[final,12pt,authoryear,times]{elsarticle}
\usepackage[utf8]{inputenc}
\usepackage{amsmath}
\usepackage{mathrsfs}
\usepackage{graphicx}
\usepackage{epstopdf}
\usepackage{enumitem}
\usepackage{siunitx}
\usepackage{float}
\usepackage{mathtools}
\usepackage{xcolor}
\usepackage{subcaption}
\usepackage{algorithm}
\usepackage{algorithmic}

\usepackage{textcomp} 
\usepackage{hyperref} 
\hypersetup{pdfauthor=author}
\date{May 5, 2021}
\newcommand{\comment}[1]{} 

\setlist{leftmargin=5cm}
 \usepackage{etoolbox}

\newcommand{\red}[1]{\textcolor{red}{#1}}

\makeatletter
\def\ps@pprintTitle{%
    \let\@oddhead\@empty
    \let\@evenhead\@empty
    \def\@oddfoot{\footnotesize\itshape
         {} \hfill{May 5, 2021}}%
    \let\@evenfoot\@oddfoot
    }
\makeatother

\begin{document}

\title{Stick-slip phenomena and Schallamach waves captured using reversible cohesive elements}

\author[cee,epfl,ULB]{Evelyne Ringoot}
\author[epfl]{Thibault Roch}
\author[epfl]{Jean-François Molinari}
\author[ULB]{ Thierry J Massart}
\author[cee,me]{Tal Cohen\corref{cor1}}
\ead{talco@mit.edu}

\address[cee]{Department of Civil and Environmental Engineering, Massachusetts Institute of Technology, Cambridge, MA 02139}
\address[me]{Department of Mechanical Engineering, Massachusetts Institute of Technology, Cambridge, MA 02139}
\address[epfl]{Civil Engineering Institute, Materials Science and Engineering Institute, Ecole Polytechnique Fédérale de Lausanne, Station 18, CH-1015 Lausanne, Switzerland}
\address[ULB]{Building, Architecture and Town Planning (BATir), Université Libre de Bruxelles (ULB), 1050 Brussels, Belgium}

\cortext[cor1]{Corresponding author.}

\begin{abstract}
Reversibility is of paramount importance in the correct representation of surface peeling in various physical settings, ranging from motility in nature, to gripping devices in robotic applications, and even to sliding of tectonic plates. Modeling the detachment-reattachment sequence, known as stick-slip, imposes several challenges in a continuum framework.
Here we exploit customized reversible cohesive elements in a hybrid finite element model that can handle occurrence of snap-through instabilities.
The simulations capture various  peeling phenomena that emerge in experimental observations,  where  layers are pulled from a flat, rigid substrate in the direction parallel to the surface. For long layers, periodicity in reattachment is shown to develop and is linked to the concept of Schallamach waves. Further, the connection between surface properties and stick-slip behavior is investigated: {we find} that stick-slip is linked to the propensity of the interface to localize deformation and damage. 
Beyond elucidating  the various  peeling behaviors and the detachment modes, the {computational} framework developed here provides a straightforward approach for investigation of complex delamination processes, which can guide the development of future applications across different scales and in various settings.

\end{abstract}
\maketitle
\section{Introduction}
\noindent Failure of bonded interfaces  is ubiquitous at different scales and in various settings. While, in the mundane, failure of an adhesive layer is a common nuisance, in advanced engineering applications, controlling this failure can lead to desired functionalities and novel fabrication methods. Several such applications are inspired by the superior performance of adhesive interfaces that are used for locomotion in the animal kingdom, with examples ranging from the scale of a single cell, to insects and lizards \citep{naturefrictionfracture, surfacesurfacesliptheoryandexperiments, reviewgeckomobility}.  At even larger scales, relative motion between tectonic plates, and the resulting seismic waves, are also triggered by interfacial failure \citep{ seismichydrogellink2, eartquakesandstickslip2, surfacesurfacesliptheoryandexperiments}. 
In certain instances, depending on the nature of the bonding interaction and the loading state, local re-bonding can occur if the two faces of the interface come back into contact,  before complete failure.  
Such reversible bonding may emerge, for example, due to interlocking of asperities on rough surfaces, or by molecular interactions, such as van der Waals forces, and can significantly alter the observed phenomena \citep{nosonovsky2007multiscale}\footnote{Note that among several mechanisms that contribute to friction \citep{nosonovsky2007multiscale}, such as plastic deformation at the interface and wear or contamination particles between the surfaces, we restrict our attention in this work to   reversible mechanisms.}.

Reversibility is a desired feature in robotics applications, where the exceptional load bearing capacities of modern adhesive layers that do not damage the climbing surface \citep{roboticsreviewnew,roboticsreview, robotics}, has facilitated the development of inspection devices for dangerous environments \citep{firstclimbingrobot, robotics}, and of high-precision soft adhesive grippers that allow manipulation of fragile objects without leaving residue \citep{grippersreview, adhesivegripper2,  medicalgripper4, adhesivegripper3}; which is a particularly useful functionality for  minimally invasive surgery \citep{ surgicalmicromanipulation, mis_gripper}. 
{Such high-load bearing capacities with unlimited cycles of detachment and reattachment, at high speeds, and without damaging even on rough surfaces have been repeatedly observed in nature.
Several competing theories have been proposed to  delineate the various characteristics of natural adhesives that allow them to exhibit such unrivaled reversibility.  The role of hierarchical fibrillar structures that are observed in several species has been proposed as a possible explanation  \citep{  adhesivemimickinggecko, geckoangle,femprestress, reviewhiearchicaladhesiongecko, hiearchicalreview, roboticsreview, yetanotherreviewfibrils, adhesivereview}. However,  it has been shown that the high load bearing capabilities observed across species \citep{labonte2016extreme} can be matched without hierarchical features, by tuning the in-plane compliance of the layer \citep{scalingthoery}. This was achieved with adhesive systems composed of a soft adhesive layer with a stiff backing \citep{scalingreview, compliance2, samearticle}. In this work we focus our attention to such bi-layer systems.}

{A recent study has shown that bi-layer adhesives can exhibit various distinct  failure modes, including the formation of an interfacial cavity near the pulling end, or the propagation of a peeling front from the opposite end, which is referred to as curling, as it is associated with bending deformation of the far region \citep{cohen_main}. Experimental observations show that this curling mode spontaneously leads to complete failure, while additional pulling is required to arrive at complete failure after the first formation of an interfacial cavity. A theoretical model that accounts for finite stiffness of the adhesive bond explains the onset of different failure modes. However, it was unable to capture the propagation of failure nor to determine the load bearing capacity. In this work we consider a similar problem setting to study the response of bi-layer adhesives to pulling along the direction of the substrate, as in a zero-degree peeling test.
In this configuration, the layer may be more prone to reattachment, which can have a significant influence on the final failure mode and the load bearing capacity\footnote{It should be noted that studies on single-layer adhesives subjected to similar loading conditions \citep{puresheartheory, compliancenewest, shearlag2} show a different failure response. This is explained by the effect of the stiffer backing on the distribution of the shear deformation, also referred to as `shear lag'.}.
}

While full control of the delamination process of soft adhesives is essential in robotics applications {and has been observed in nature}, successive de-bonding and re-bonding can occur spontaneously within regions of a loaded 
interface. Such phenomena have been referred to as \textit{stick-slip} events \citep{brace1966stick}, \textit{Schallamach waves} \citep{SCHALLAMACH}, and \textit{self-healing pulses} or \textit{Heaton waves} \citep{HEATON}, and have long been established as a prominent mechanism in earthquakes \citep{ seismichydrogellink2, eartquakesandstickslip2, surfacesurfacesliptheoryandexperiments}. {The driving mechanism is similar for both the formation of Schallamach waves in soft materials \citep{SCHALLAMACH, fea_schallamach} and slip pulses in geologic faults \citep{slippulsestensioncompression}: local buckling caused by tension-compression fields drives the formation and propagation.}

{The apparent analogies between stick-slip processes in tectonic plate movement, and  debonding of soft adhesive layers, have motivated  several studies that consider the latter as a desktop scale representation of the former \citep{seismichydrogellink2, surfacesurfacesliptheoryandexperiments, seismicpicture}. 
For example, waves that emerge  during  sliding  contact  between  a hard indentor and a soft substrate have been considered in several experimental and theoretical studies 
\citep{femspheres, spheresphereanlytical2, schallamachwaves2006,analyticalsurfacesurface, fea_schallamach, sphereslidingc, sphereslidinga, sticksliptheory, numericaldifficultiesmathstickslip,spoolslidingb,  {longarticlestickslipsoft,seperationpulseslippulse}, spoolslidinga, sphereslidingb}.
Finite element models used in these studies describe the development of the detachment and  the formation of the first Schallamach wave, but they do not include reattachment. Recently, continuum models including friction-adhesion contact coupling have been proposed to treat this challenge \citep{contactreversibility, interfacemodels, adhesivecomputing, adhesivefriction2, frictionaladhesionfem5} and have even been applied to sliding processes \citep{femcoupledadhesionfriction}.  Despite the recent advances,  
literature on theoretical models that can capture the entire, unsteady, process of delamination,  in presence of stick-slip events, is in a nascent state. Addressing this limitation may thus provide additional insights into various phenomena and may 
even help to explain why adhesion based motility exhibited in nature has yet to be matched by its synthetic counterparts. }

{
Capturing the propagation of failure from its initiation to complete detachment in a theoretical model is especially challenging when considering  zero-degree loading. This  complexity is primarily due to the limited available computational tools that can account for the layers ability to re-attach, at a new location, if it comes back into contact with the substrate. Nonetheless, experimental evidence of this phenomenon has been repeatedly reported \citep{ oldinterfaceslip, selfhealingshearpulse,  stickslipzerodegree, slidingimportant,  puresheartheory, cohen_main, debondingslip, newstudypeelinginstability}, and can occur even when pulling at a prescribed angle \citep{slidingimportant, stickslipevidenceexperimental2}. The stick-slip behavior is also evident from the force-displacement curves that show sudden drops in the load followed by recovery  in controlled pulling tests  \citep{puresheartheory, cohen_main}.}

To the best of our knowledge, to date, no  theoretical model has provided a comprehensive account of the various phenomena that emerge when  reattachment of an adhesive interface is possible, as observed in zero-degree peeling of bi-layer adhesives. Understanding the range of model parameters for which such phenomena occur and their influence on the global behavior of the adhesive pad, can pave the way to engineering of advanced adhesive systems, and can elucidate phenomena observed in the natural world at various scales. 

In a finite element framework, cohesive elements are a natural choice for representing interfacial forces that are weaker than the  bonds in the bulk of the  material, {where} the failure is localized to the interface. 
Other than capturing fracture phenomena in stiff or brittle materials, such as metals and concrete
\citep{cohesivesteel,quasibrittlereview}, cohesive elements are commonly used  for modeling soft adhesives \citep{cohesivefempolymerpad, fempeelinganglesinglelayer, cohesivezonepsa}.
However, most cohesive laws do not include healing or reattachment. 
Continuum damage models have been proposed to model healing by allowing  broken elements to reverse damage and have been successfully implemented in finite element analysis using discrete element methods and mesh-free methods \citep{continuumhealingreview, review2020healing}. To determine the healing kinetics, recent studies have included, for example, chemo-mechanical coupling \citep{ chemomech4,recentchemmechcoupling, chemmechrecent2, chemomech3}, thermodynamic healing \citep{thermodynamichealing1, thermodynamichealing1bis} and biological factors \citep{biologicalhealingcontinuumdamage}. 
However, this approach is not well adapted to capture interfacial phenomena considered here. 

{The objective of this work, is to capture reversible interfacial peeling phenomena  that may emerge in various delamination processes, including instability and stick-slip events. Hence, in this study, we propose a modification to the classical cohesive law by \cite{ortizpandolfi}, to allow full recovery of a cohesive bond upon contact of the two sides of the interface. We will apply  this modified cohesive law  to study   peeling phenomena in elastic bi-layer adhesives that are capable of large deformations, as they are pulled along the direction of the substrate.} 

This manuscript is organized as follows: the next section describes the physical problem setting with all relevant model parameters, and provides an overview of various physical phenomena that emerge in experimental observations of this system. Modeling of the interface response is documented in Section \ref{sect:interfaceprop}. The specialized finite element algorithm developed to capture the peeling response is detailed in Section \ref{sect:procedureology}.  Next, Section \ref{sect:results} presents the simulation results {and their discussion}, showing agreement with our experimental observations. We further extend the analysis to capture the response of infinitely long layers and to elucidate the constitutive sensitivities.  Finally, concluding remarks are given in Section \ref{section:discussion}.

\section{Problem setting and observations of peeling response} \label{sect:problemstatement}

\begin{figure}[htp]
\centering
\includegraphics[width=0.5\linewidth]{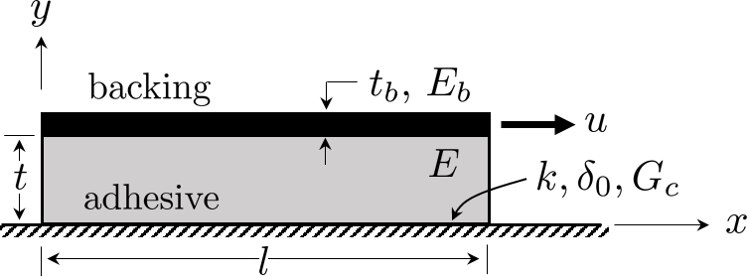}
\caption{Illustration of problem setting. A bi-layer adhesive pad is composed of a soft adhesive and a stiffer backing. The adhesive surface is placed on rigid substrate and pulled along the horizontal coordinate, $x$, by controlling the displacement, $u$. Shown on the Figure are the stiffness of adhesive and backing $E$ and $E_b$, respectively, the properties of the adhesive surface (i.e. bond stiffness $k$, normalized opening at initiation of the softening phase $\delta_0$, and surface energy\protect\footnotemark $G_c$), and the geometric dimensions (i.e. the length of the layer $l$,  the thicknesses of backing $t_b$, and the adhesive layer $t$)}
\label{fig:prop3}
\end{figure}

Consider an elastic bi-layer composed of an adhesive layer of length 
$l$ and thickness $t$ that is perfectly attached to a stiffer backing of thickness $t_b$ and placed on a smooth, infinitely stiff substrate, as illustrated in Figure \ref{fig:prop3}. Restricting our attention to plane-strain deformation, we define the Lagrangian coordinates $(x,y)$ such that the bi-layer adhesive occupies the region \begin{equation} 0\leq x \leq l, \quad 0\leq y\leq t+t_b. \end{equation}

A horizontal displacement, $u$, is applied to the backing at $x=l$ - henceforth referred to as the \textit{pulling end}, and is associated with a resultant force $F$ (per unit length).

We assume that both the adhesive layer and the backing of this bi-layer system are incompressible and capable of large elastic deformations that are well described by the neo-Hookean hyperelastic model. Their strain energy density can thus be written in terms of a single constitutive parameter, the elastic modulus $\hat E$, in the form $W=\hat E(I_1-3)/6$ where $I_1$ is the first invariant of the right Cauchy-Green deformation tensor. We denote the distinct elastic moduli of the adhesive layer and the backing by $\hat E=E$ and $\hat E=E_b$, respectively. \footnotetext{ The energy necessary to detach a unit area of the adhesive surface from its substrate is referred to here as `surface energy', and is equivalent to the term `fracture energy', which is commonly used for cracks.}

Before defining the properties of the interface, it is instructive to examine, in more detail, the different modes of failure that may emerge in this system. Hence, we conduct a series of observations that we describe next. We emphasize that while earlier studies have provided a more comprehensive experimental investigation of the considered bi-layer system \citep{cohen_main}, 
we provide these observations here  as visual context: for choice of a specific cohesive law; for qualitative comparison; and for clarification of numerical results\footnote{We emphasise that these observations are not intended as a comprehensive experimental investigation.}.

\begin{figure}[htp]
\centering
\includegraphics[width=1\linewidth]{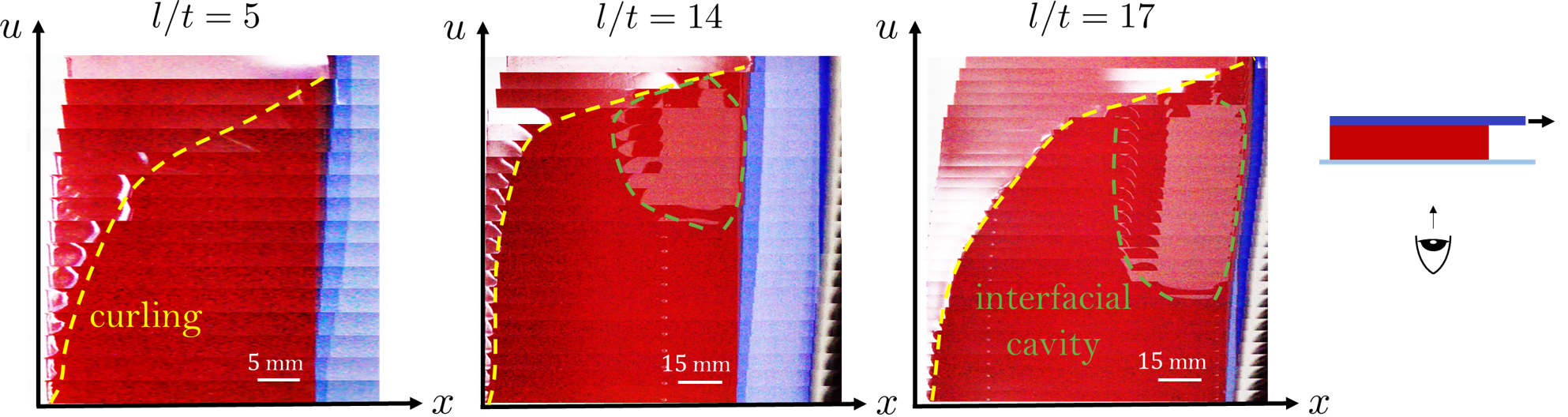}
\caption{Three representative observations of the peeling process shown as a sequence of snapshots for increasing displacement $u$, along the vertical axis. (The vertically stacked rectangles show the configurations of the pad for an increasing displacement.) Corresponding videos can be found in \textcolor{blue}{\href{https://www.dropbox.com/s/wrsnip2p6lwm5rv/combined_video.mp4?dl=0}{supplementary material}}. To obtain a 2D interpretation of the process (some 3D effects can be observed in the videos), only a cropped region in the mid section along the width of the layer is included for each displacement. The length of the layer is shown along the horizontal coordinate - $x$.   Observations are made from below, through a glass substrate. Peeled regions appear brighter due to changes in the refractive index. The propagation of a curling front is marked by a dashed yellow line.  Regions of interfacial cavity are marked by a dashed green line. For all layers $t=6.25$ mm, $t_b=4$ mm, and the out of plane width is $30$ mm. The elastic moduli are $E=0.7$ MPa, and $E_b=20$ MPa. For additional details on sample fabrication and experimentation see {\ref{app:fabrication} }.}
\label{exp}
\end{figure}

The different modes of failure and the possible influence of re-bonding is demonstrated by three observations in Figure \ref{exp}. For all cases we use   Polydimethylsiloxane (PDMS) to fabricate both the backing and the adhesive layer with thicknesses $t_b=4$ mm and $t=6.25$ mm, and with elastic moduli $E_b=20$ MPa, and $E_b=0.7$ MPa, respectively. A pigment is used to distinguish between the adhesive (red) and the backing (blue). The only difference between the three bi-layer systems is their length, as shown by the different values of $l/t=5,14,17$. Observations are made by looking through the glass substrate as the backing is pulled using a mechanical testing machine. Additional details on the fabrication process and testing can be found in { \ref{app:fabrication}}. Videos of the peeling are provided in the \textcolor{blue}{\href{https://www.dropbox.com/s/wrsnip2p6lwm5rv/combined_video.mp4?dl=0}{supplementary material}}. 

First, for the shortest layer $(l/t=5)$, we observe that peeling initiates from the far end (i.e $x=0$) and propagates towards the pulling end, up to complete failure. In comparison with the pulling rate, which is effectively shown by the slope of the line that divides the red and blue regions in the figure, the curling front propagates faster, and accelerates as pulling progresses. In contrast,  peeling of the layer with $l/t=14$ propagates differently. More displacement of the pulling end is needed to initiate the propagation of the curling front (notice the different scale bars indicated for the different cases), which occurs simultaneously with the formation of an interfacial cavity. Shortly after its formation, the interfacial cavity is shown to split into two regions while an intermediate region appears to re-bond. This \textit{shedding} event\footnote{As shown in the video, the main cavity releases a smaller cavity, which due to re-bonding, appears to propagate toward the far end.  Hence the use of the term shedding.} seems to allow the main cavity (nearest the pulling end) to maintain a nearly constant size. Finally, as peeling progresses, the curling front rapidly engulfs as it propagates towards the pulling end. Quite interestingly, rather than merging the debonded regions, the interfacial cavity reattaches to the substrate as the curling front approaches. 
For the longest layer, with $l/t=17$, the nearly constant size of the main cavity and its closure in response to the approaching curling front, becomes even more pronounced.

In all cases, $3D$ effects are observed upon initiation of failure and during its propagation. In particular, we find that the interfacial cavity nucleates from defects near the edges and then propagates to the entire width.  These effects are not accounted for under the plane-strain assumption for which the entire length of the cavity would form simultaneously. It is  thus expected that the model predictions, which we describe next, will result in a tougher response with higher load bearing capacity. 

\section{Interface properties and reversible cohesive elements} \label{sect:interfaceprop}

From our observations in the previous section, re-bonding is clearly shown to play a key role in the peeling process. 
Moreover, as seen from the large displacements that are attained before complete failure, it can occur at a distance from  the initial bonding location. To understand the influence of re-bonding on the peeling process and on the load bearing capacity of the layer, we describe next a cohesive law that can capture peeling and re-bonding at a new location, upon contact between the two surfaces. 

\begin{figure}[htp]
    \centering
    \includegraphics[width=0.7\textwidth]{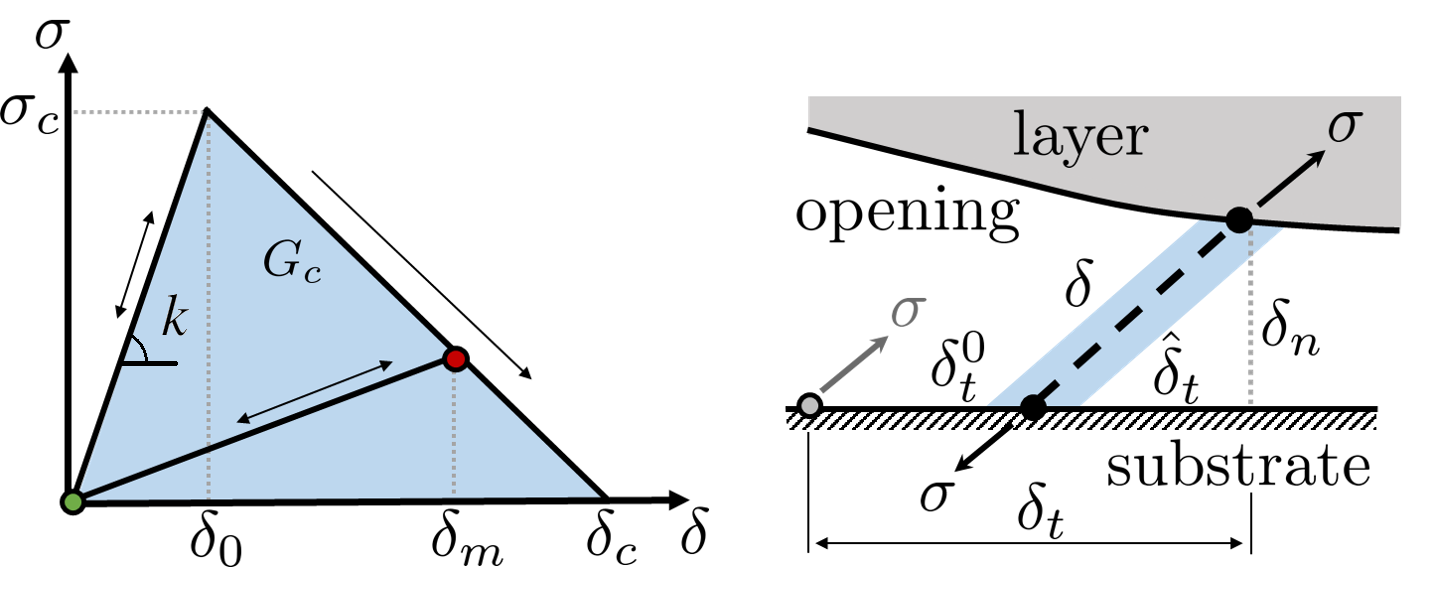}
    \caption{Bi-linear cohesive law (left) and illustration of corresponding parameters (right). The stress in the cohesive zone is a function of the opening $\delta=(\delta_n^2+{\hat\delta_t}^2)^{1/2}$ and is oriented along the opening direction, where  the tangential displacement with respect to the unloaded  state of the bond is $\hat \delta_t =\delta_t - {\delta_t^0}$ and the normal opening is $\delta_n$. If unloading initiates on the unstable branch (i.e. $\delta_0<\delta<\delta_c$), as indicated for example by the red circle, then the response follows a softer linear decay which intercepts the origin (indicated by the green circle). If the element breaks and then re-adheres at a new location (i.e. $\delta_t^0\neq 0$), the opening and the corresponding direction of the interface stress are adjusted accordingly. }
    \label{fig3}
\end{figure}

While several forms of a cohesive law can be conceived, without evidence of a specific physical form, we choose here the simplest possible law;  a bi-linear stress-opening relationship\footnote{Note that for consistency,  we describe the cohesive element in terms of its stress-opening response, where the stress is the force per unit area of the cohesive element and the opening is the gap between top and bottom surface of the cohesive element.}, as illustrated schematically in Figure \ref{fig3}. Here $\delta$ represents the absolute distance between two corresponding segments on either side of the adhesive  interface,
hence the normal and tangential components of the opening affect the cohesive response with the same weight, 
and the \textit{interface stress}, $\sigma$, acts along the direction of the opening ($\delta$).  Initially, the bonds deform elastically with  stiffness $k$ (per unit area). Then, once an opening of $\delta=\delta_0$ and the corresponding critical stress $\sigma=\sigma_c=k\delta_0$ are reached, an{ unstable softening response } is activated up to full detachment at $\delta=\delta_c$ where $\sigma=0$. {The total energy expended is $G_c=\sigma_c\delta_c/2$, as indicated by the shaded blue region.}  Overall, this cohesive response can be represented by three independent material parameters. Investigation of the constitutive  sensitivity in the following sections,   will center on the set ($k$, $\sigma_c$, $G_c$), or equivalently, its dimensionless counterpart (as described in Section \ref{sect:results}).

\smallskip
\noindent \textbf{Unloading response.}  Before describing the re-attachment of the adhesive bond, we first consider situations in which unloading initiates before complete  bond breakage is achieved. On the stable branch, unloading would merely reverse the direction of the response along the stress-opening curve. However, if unloading occurs on the unstable branch, the response  must depart from the original bi-linear law, resulting in hysteresis. The response then follows   a new bi-linear form with the point of departure  from the unstable branch (indicated by the red circle on Figure \ref{fig3}) being the new peak stress at $\delta_m$, as in the classical formulation of the Camacho–Ortiz linear irreversible cohesive law \citep{CAMACHOortiz}. Mathematically, the interface stress can be written as
\begin{equation}\label{sigma}
    \sigma=\sigma_c\begin{cases} {\delta}/{\delta_0} & 0\leq\delta,\delta_m\leq \delta_0    \\ \frac{\delta}{\delta_m}\left(\frac{\delta_m-\delta_c}{\delta_0-\delta_c} \right) & 0\leq\delta\leq\delta_m, \quad \delta_0\leq\delta_m\leq\delta_c\\ 0 & \delta_m\geq  \delta_c \end{cases} 
\end{equation}

Additionally, we will examine the separate contributions of the tangential displacement $\delta_t$ and the normal opening $\delta_n$, which define the absolute opening through the equality \begin{equation}\delta=\sqrt{\delta_n^2+(\delta_t-\delta_t^0)^2}, \end{equation}where $\delta_t^0$, accounts for an initial gap between the corresponding segments on either side of the interface, in the adhered state. Before re-bonding occurs $\delta_t^0\equiv 0$.  Nonetheless, if the bond breaks and then re-bonding occurs, this gap can change, as will be explained shortly.   

The corresponding tangential and normal components of the interface stress are \begin{equation}\label{stsn}
    \sigma_t=\sigma\frac{\delta_t-\delta_t^0}{\delta}, \quad \text{and} \quad \sigma_n=\sigma\frac{\delta_n}{\delta},
\end{equation} respectively. Note, that the above formulation of the cohesive law in equations \eqref{sigma}-\eqref{stsn}, assumes no interpenetration, i.e. $\delta_n\geq0$. However, if $\delta_n\leq0$ then an impenetrability condition is enforced.

\smallskip
\noindent\textbf{Contact penalty.} Considering impenetrable substrates, we include a numerical penalty for interpenetration, which is activated if $\delta_n\leq0$. Then, the tangential and normal component of the  interface stress are calculated separately as
\begin{equation}
      \text{if } \quad \delta_n\leq0: 
      \qquad \sigma=\sigma_n=k_p \delta_n, \quad \sigma_t=\sigma(\delta_t-\delta_t^0), 
\end{equation}
where the stiffness $k_p$ is a numerical interpenetration penalty.

\smallskip
\noindent\textbf{Damage accumulation.} 
For interpretation of the numerical results, it is instructive to determine the level of damage within the cohesive zone.  Hence, we define a damage parameter $D$ that varies linearly between $0$ and $1$ as damage accumulates, up to complete failure 
\begin{equation}
    D=\begin{cases} 0 & 0\leq\delta,\delta_m\leq \delta_0,    \\ \frac{\delta_m-\delta_0}{\delta_c-\delta_0} & 0\leq\delta\leq\delta_m, \quad \delta_0\leq\delta_m\leq\delta_c,\\ 1 & \delta_m\geq  \delta_c. \end{cases} 
    \label{eq:damage}
\end{equation}
Although in conventional cohesive laws, damage is mechanically irreversible, in this work, we consider situations in which renewed contact between the two faces of the interface allows to recover the adhesive bond, and as such, to reverse the damage. 

\smallskip
\noindent\textbf{Reversibility.}  Finally, we describe the key feature of the cohesive elements used in this work, their reversibility. For simplicity, we will assume here that if a segment of the adhesive surface reestablishes contact with the substrate, then the bond is fully recovered. Nonetheless, it would be rather straight forward to relax this assumption to account for a deterioration of the adhesive capability from one delamination event to another, or also its dependence on the peak value of the  normal compressive interface stress $\sigma_n$ applied in reestablishing the contact. 
Considering situations in which large displacements can lead to re-bonding at a new location that may be distant from the initial bond, it  seems natural to allow for new segments on either side of the interface to form bonds. However, such an approach is mathematically complex and computationally very costly. Instead, in this work, we   implement the re-attachment between the initially linked pair, while re-positioning the interface stress to  correctly account for the location of reattachment. An artifact of this approach is that the reaction interface stress acting on the substrate will be located at the position of the initial bond (Figure {\ref{fig3}}). By limiting our attention to peeling from flat rigid substrates, this has no bearing on the results.
Mathematically, the implementation of this simplified approach is straightforward. Following the elongation and breakage of the initial bond, if a segment of the adhesive surface reestablishes contact with the substrate, such that $\delta_n\leq 0$, then the full cohesive energy $(G_c)$ is recovered at the new tangential location $\delta^0_{t}$ and the damage is set back to $D=0$, namely
\begin{equation}
    \text{if} \quad D=1\quad \land \quad \delta_n\leq0 \quad \text{then set}\quad D=0, \quad \delta^0_{t}=\delta_t.
\end{equation}
Then, as the bond is further deformed, the tangential displacement is measured from this point. 
In this framework, re-attachment events can happen multiple times and it is sufficient to declare a different $\delta^0_{t}$. Accordingly, at the onset, $\delta_t^0\equiv 0$ for all $x$. 

In this section we have established a framework that describes the response of reversible cohesive elements. For simplicity, this framework considers a bi-linear response, it is limited to adhesion on flat rigid substrates, and considers full recovery of the adhesive bond upon contact with the substrate. {Nonetheless, it is straight forward to extend this framework and relax these assumptions to account for more complex, rate-dependent, adhesive response and mixed mode failure \citep{snozzimolinari,wang2014crack}.} 
In the next section we provide the details of the numerical exploitation of these cohesive elements by adapting a  readily available finite element framework, and we describe the solution procedure.

\newcommand{\mypar}[1]{\smallskip
\noindent 
\textbf{#1} }

\section{Numerical Implementation}\label{sect:procedureology} 
The considered problem encompasses large deformations, both in the bulk and at the interface, as well as local unstable response of interface bonds and their reversibility; it is thus highly nonlinear and requires  specialized  numerical implementation. The ingredients that are included in the numerical scheme to resolve these nonlinearities, and to minimize computation times, are detailed here:

\mypar{Setup of the simulation.} The problem setting and constitutive properties described in Section \ref{sect:problemstatement} are implemented numerically to investigate the peeling response. 
The pad is loaded by imposing an increasing displacement $u$ of the edge of the backing (i.e. at $x=l$ and $y\in[t,t+t_b]$), while not allowing vertical displacement along $y$ of this edge. Zero thickness, reversible cohesive elements, with a response described in the previous section, are included at the bottom of the pad (i.e. at $y=0$ and $x\in[0,l]$). At the interface between the adhesive layer, a rigid substrate is included in the simulation to ensure proper boundary conditions. The remaining free surfaces of the layer remain unloaded. 

\comment{
\red{I think we can remove the equations and thus avoid new notations OK
\begin{equation}
    \forall x=l, \quad t \leq y \leq t+t_b : \qquad \Delta y=0,\quad \Delta x= u
\end{equation}
The respective vertical and horizontal displacement of the edges are designated $\Delta y$ and $\Delta x$.
The other edges of the surface are left free:
\begin{equation}
    \forall ( x=0, \; 0 \leq y \leq t+t_b) \; \lor \;
    ( x=l, \; 0 \leq y \leq t) \; \lor \;
    (y=t_b, \; 0 \leq x \leq l): \quad
    \mathbf{f}^{ext} = \mathbf{0}
\end{equation}
The vector of externally applied forces if designated $\mathbf{f}^{ext}$.
On the bottom of the pad, cohesive elements are used to represent surface attachment at $x=0$. The cohesive elements are adapted to include reattachment as described in the previous section.
To ensure proper boundary conditions, an elastic material is included below the surface, the displacement \red{of which} is fully blocked:
\begin{equation}
    \forall y \leq 0: \Delta y=0,\quad \Delta x= 0
\end{equation}}}

\mypar{Meshing and calculation software.} The mesh is generated using the open-source mesh-generation software Gmsh \citep{Gmsh} and the simulations are performed with the open source multi-core finite element solver, Akantu \citep{akantu1, akantu2}. The bulk materials are meshed into T3 elements with linear interpolation and the adhesive surface is meshed into zero thickness cohesive  4-node elements with a Gauss integration scheme. The spatial discretization is chosen such that the cohesive process zone is sufficiently discretized. Visual  post-processing is achieved using the open-source visualization application ParaView \citep{paraview1, paraview2}. To avoid element locking in the simulations, we use the modified neo-Hookean law to allow for small levels of compressibility. Numerical values used in the simulation are provided in \ref{app:numericalparams}.

\mypar{Hybrid solution procedure.} 
In the finite element framework, the static problem can be described by the system of equations

\begin{equation}
    \mathbf{f}^{int}(\mathbf{u})= \mathbf{f}^{ext} 
    \label{eq:staticsol}
\end{equation}
Using its linearization at each iteration of a Newton-Raphson-based scheme, a linear system of equations is solved at every iteration
\begin{equation}
    \mathbf{K \; \delta u} = \mathbf{f}^{ext}  - \mathbf{f}^{int}_{previous} 
\end{equation}
where $ \mathbf{K}$ is the tangent stiffness of the structure that depends on $\mathbf{u}$, the set of nodal displacements\footnote{To avoid confusion, note the difference between $\mathbf{u}$ and the applied displacement $u$}.
 Solving this system using a static solution procedure incurs unavoidable convergence issues that result from the  cohesive elements and the nonlinear material laws. To treat such issues, the solution procedure can employ  increasingly smaller displacement steps $\Delta u$.   However, this requires longer computational times and, moreover, once interfacial bonds begin to break,   convergence may not be achieved. 
 
 Alternatively, one can employ an explicit dynamic  solver, formulated using the central difference solution procedure 
\begin{equation}\label{eq:dyn_solv}
\mathbf{M}\mathbf{\ddot u(\theta)} + \mathbf{K}\mathbf{u(\theta)} = \mathbf{f}^{ext}(\theta)
\end{equation}
where nodal accelerations $\mathbf{\ddot u(\theta)}$ and velocities $\mathbf{\dot u(\theta)}$ are permitted, and $\mathbf{M}$ is the lumped mass matrix. In this time stepping scheme the nodal forces, displacements, accelerations and velocities depend on time, $\theta$, and integration is performed using time increments, $\Delta \theta$, which correspond to displacement steps $\Delta u=v\Delta\theta$, where $v$ is the pulling rate. {To  capture a `nearly' quasi-static peeling response (i.e. with negligible levels of inertia), the choice of a specific pulling rate is a compromise between achieving minimal inertial effects}, and allowing for reasonable simulation times\footnote{In this work numerical stability is achieved for  $\Delta \theta$ of  the order of the {critical time step of the bulk}, $0.1 \mu s$, resulting in long computation times of $\sim 6$ hours on $28$ {cores}. The maximal critical time step is defined as the time the fastest wave needs to travel the characteristic length of the mesh\citep{akantu} in the bulk.}. 

Overall, both the explicit and implicit solvers, represented by equations \eqref{eq:staticsol} and \eqref{eq:dyn_solv}, respectively, have significant shortcomings. Nonetheless, the static solution  performs well before onset of bond  breakage 
and the dynamic solution procedure, although it is time intensive, allows to handle instabilities  that may emerge in the peeling process.  To take advantage of both of these methods, we use in this work a hybrid approach: a static procedure is employed until the first section along the surfaces reaches a damage value of $D=0.1$, at which point the dynamic solution procedure is initiated. { The dynamic solution procedure is not meant to capture  inertial effects; rather it is deployed  to capture the non-equilibrium transition between   static equilibrium states, it is accordingly damped to the degree of becoming nearly quasi-static.}  
To avoid imposing a numerical shock in the loading procedure, at the transition into the dynamic scheme, the pulling velocity $v$ is gradually increased up to its target value.

\mypar{Damping.} In this work we aim to capture the quasi-static peeling response of the system {by sufficiently damping the dynamic response, in an approach similar to \citet{rice1993damping}}. Without justification for a specific physical  damping mechanism, we employ artificial damping by imposing a numerical correction factor on the velocity. Accordingly, at every time step the equation of motion \eqref{eq:dyn_solv} is solved, and the predicted velocity is reduced by application of the factor, $c<1$, such that
\begin{equation}
\mathbf{\dot u (\theta)}= c \cdot \mathbf{\dot u}^{pred} (\theta) 
\label{eq:dynamicsol}
\end{equation}
\comment{
\red{\begin{equation}
   \text{we do not need the equation again: ok }\qquad \mathbf{M}\mathbf{\ddot u(\theta)} + \mathbf{K}\mathbf{u(\theta)} = \mathbf{f}^{ext}(\theta) \qquad 
\mathbf{\dot u (\theta)}= c \cdot \mathbf{\dot u}^{pred} (\theta) 
\label{eq:dynamicsol}
\end{equation}}
}
The initial velocity before damping correction $\mathbf{\dot u}^{pred} (\theta)$ is calculated according to the standard explicit central difference method.
\begin{equation}
    \mathbf{\dot u}^{pred}(\theta)=\mathbf{\dot u}(\theta-\Delta \theta) + \frac{\Delta \theta}{2} (\mathbf{\ddot u}(\theta)+\mathbf{\ddot u}(\theta-\Delta \theta))\\
\end{equation}

The damping factor $c<1$ is defined based on the time step-independent constant\footnote{This constant is calculated using the formula $\theta_{1\%}(c)={\Delta \theta \ln(c)}/ {\ln(0.99)}$ to eliminate dependence on the time step $\Delta \theta$. } $\theta_{1\%}(c)$.
The damping constant is a numerical parameter and can in practice be tuned to achieve desired levels of damping. In this work we seek an optimal value for $\theta_{1\%}$ that minimally influences the force-displacement response while eliminating vibrations. Therefore, we determine $\theta_{1\%}$ by considering a simplified benchmark problem: a block of linear elastic material of cross-section $t\times t$ is perfectly adhered to the substrate and subjected to tensile deformation by pulling its top surface at the same velocity $v$ (as in the simulations) until 1\% strain and releasing it to observe oscillations resulting from the loading relaxation. The desired value leads to critically damped behavior, such that the block returns rapidly to its stable position at rest without oscillations. 

\mypar{Snap-through.}
As observed from our experiments in Section \ref{sect:problemstatement}, failure can occur abruptly and thus,   the failure of one cohesive element can lead to a cascade of interfacial failure even without continuing to pull, namely while holding the pulling displacement $u$ constant. {This  physical snap-through instability is also encountered in experimental settings (see Fig. \ref{exp} and corresponding video).} Eventually, this rapid delamination arrests as a new quasi-static equilibrium state is found. However, if the pulling progresses while such snap-through events occur, the response can be highly dependent on the pulling rate while inertial effects become dominant and numerical damping has a non-negligible effect on the applied force.  Thus, in this work, to numerically capture a `nearly' quasi-static response, once a single cohesive element fails, the numerical procedure continues but with  constant displacement (i.e. $v=\Delta u=0$) until a new equilibrium is achieved. Then, the velocity is gradually introduced again. This procedure is continued until 95\% of the total adhesive surface has failed, avoiding the final dynamics linked to development of failure of the complete pad.

\begin{algorithm}
\caption{Solution procedure} 
\label{alg:solutionsummary}

\begin{enumerate}
    \item {\bf{Initialize the system:}} $u_0=0$, and $\mathbf{u_0},\mathbf{\dot u_0},\mathbf{\ddot u_0}=\mathbf{0}$.
    
    \item {\bf{Static solver:}} Update $u_{i}=u_{i-1}+\Delta u$. 
    Use \eqref{eq:staticsol} to calculate $\mathbf{u}(u_i)$. Use  \eqref{eq:damage} to obtain $D(x)$. 
    
    If  {$\underset{x\in [0,l]}{\rm max } \left\{ D(x) \right\}\geq0.1$}, continue to next step;   {otherwise, repeat this step.}
    
    \item {\bf{Hold dynamic solver:}} Update $u_{i}=u_{i-1}$ and  $\theta_{i}=\theta_{i-1}+\Delta \theta$. Use  \eqref{eq:dyn_solv} to obtain $\mathbf{ u}(\theta_{i})$, $\mathbf{\dot u}(\theta_{i})$ and $\mathbf{\ddot u}(\theta_{i})$. Use  \eqref{eq:damage} to obtain $D(x)$. Repeat for 10 steps.
    
    If more than 95\% of the surface failed, stop the simulation. 
    
    If $\forall x \in (0,l), \forall j \in (i-10,i-1): D_i(x)=D_j(x)$, continue to the next step;  otherwise, repeat the current step.
    
    \item {\bf{Accelerate dynamic solver:}} Increase the pulling velocity with a triangular acceleration profile over 100 $\mu s$ by varying $\Delta u$, along constant $\Delta \theta$. Update $u_{i}=u_{i-1}+ \Delta u$ and $\theta_i=\theta_{i-1}+\Delta\theta$.
    
    Calculate the state of the system $\mathbf{ u}(\theta_i)$, $\mathbf{\dot u}(\theta_i)$,$\mathbf{\ddot u}(\theta_i)$ using  \eqref{eq:dyn_solv}. Use \eqref{eq:damage} to obtain $D(x)$. 
    
    If $\exists x \in (0,l): ((D_i(x)-1) (D_{i-1}(x)-1))=0 \land (D_i(x) \neq D_{i-1}(x))$, return to {step 3}; {otherwise continue with this step.}
    
    Once the target velocity $v$ is reached, continue to the next step.
    
    \item {\bf{Steady dynamic solver:}} Update $u_{i}=u_{i-1} + \Delta u$ with a constant $\Delta u=v\Delta \theta$ and $\theta_i=\theta_{i-1}+\Delta\theta$. 
    
    Calculate the state of the system $\mathbf{ u}(\theta_i)$, $\mathbf{\dot u}(\theta_i)$,$\mathbf{\ddot u}(\theta_i)$ using  \eqref{eq:dyn_solv}. Use \eqref{eq:damage} to obtain $D(x)$. 
    
    If $\exists x \in (0,l): ((D_i(x)-1) (D_{i-1}(x)-1))=0 \land (D_i(x) \neq D_{i-1}(x))$, return to {step 3}; otherwise, repeat this step.
\end{enumerate}
    

\end{algorithm}

\newpage
\section{Results {and discussion}} \label{sect:results}
In this section we apply the {computational} model, described in the previous sections, to study the influence of reversible adhesive bonds on the peeling process. For a bi-layer adhesive pad, of given aspect ratio $l/t$, the problem is fully defined by a set of five independent dimensionless parameters 
\begin{equation}\label{dimensionless}
  \alpha = \frac{t_b}{t},\qquad
    \beta = \frac{E_bt_b}{Et}, \qquad \gamma = \frac{k t }{E}, \qquad
    \Gamma=\frac{G_c}{Et}, \qquad
    \zeta = \frac{\sigma_c}{E}.
\end{equation}
The first three of these  parameters are sufficient to describe the elastic response prior to debonding, as shown in the theoretical model by \cite{cohen_main}. Therein, assuming linearly elastic response, the deformation of the layer  was captured, thus allowing to infer the location of initiation of failure, but not its propagation. An investigation of the model sensitivities finds that curling response becomes dominant  for increasing values of the longitudinal stiffness ratio - $\beta$, for decreasing values of the dimensionless bond stiffness - $\gamma$, and for smaller aspect ratios - $l/t$ \citep{cohen_main}. However, the response is shown to be insensitive to the thickness ratio - $\alpha$. 
In this work, to capture the entire delamination process, two additional dimensionless parameters are introduced: the dimensionless surface energy - $\Gamma$, and the dimensionless load bearing capacity  - $\zeta$. 
To elucidate the role of the interface properties in  determining the peeling response, the sensitivity analysis in this work will center on the last two dimensionless parameters, while considering different aspect ratios $(l/t)$.

Next, we define the dimensionless pulling displacement, and the corresponding dimensionless applied force as 
\begin{equation}
    \Delta=\frac{u}{t}, \qquad f=\frac{F}{Et},
\end{equation}
respectively. Recall that $F$ is defined as the force applied per unit depth of the layer.

\subsection{Qualitative agreement with observations}
To examine the ability of our model to capture the different modes of failure that have been observed in Figure \ref{exp}, we first consider  layers with 
\begin{equation}\label{base_set}
\alpha=0.2,\qquad \beta=12,\qquad \gamma=2.56,\qquad \Gamma=5\cdot 10^{-4},\qquad\zeta=0.032,
\end{equation}  
Note that this base set of parameters will be used in all simulations, unless noted otherwise. {The dimensional values implemented in the simulations are provided in { \ref{app:numericalparams}}.}

\begin{figure}[ht]
\centering
\includegraphics[width=0.99\linewidth]{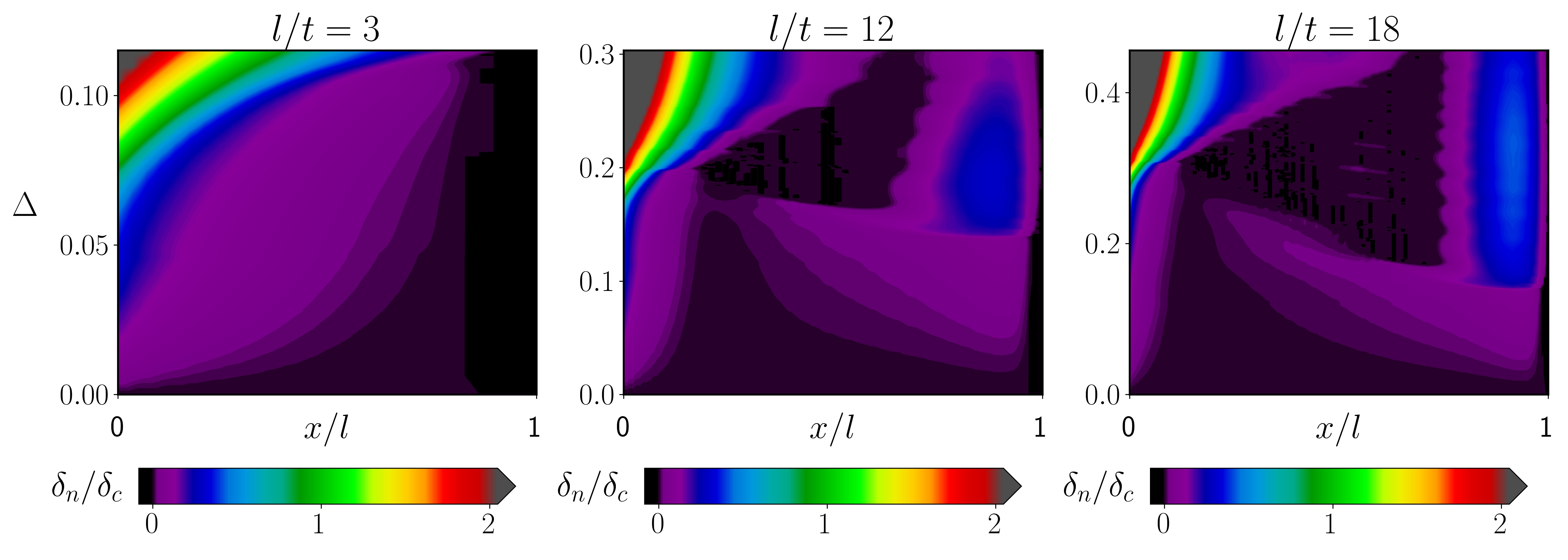}
\caption{Maps of normal opening of the interface for parameter values in \eqref{base_set} and for $l/t=3,12,18$, analogous to the observations in Figure \ref{exp}. On the y-axis the imposed displacement $\Delta=u/t$ is increased, and on the x-axis the pad length $x/l$ is displayed. For the $l/t=3$, peeling is initiated  at the opposite end $(x/l=0)$, as is typical for the curling response. For the larger $l/t$ values, an interfacial cavity develops first near the pulling end $(x/l=1)$ and curling is triggered at larger displacements leading to complete failure.
}
\label{fig:qualit}
\end{figure}

To visually compare the numerical results with the observations, we first show maps of the dimensionless normal  displacements, $\delta_n/\delta_0$, on the $\Delta,x/l$ plane for three different aspect ratios $l/t$, in Figure \ref{fig:qualit}. Note that these results, in contrast to the observations in Figure \ref{exp}, are shown in the Lagrangian frame (i.e  the $x/l$ locations refer to the undeformed state). Additionally, while in the observations  only two shades are observed and correspond to regions that are in contact (dark) or removed (bright) from the substrate, in the numerical results the lift off of the substrate is quantified. Nonetheless, the qualitative agreement between the observed phenomena and the simulation is apparent. For the shorter layer, lift off begins at the far end and is shown to accelerate as it approaches the pulling end. For longer layers (i.e. larger aspect ratio $l/t$) the formation of an interfacial cavity is clearly observed. Furthermore, it appears to maintain a nearly constant size as pulling progresses and regions in its wake (i.e. at smaller values $x/l$) come back into contact with the substrate. Finally, curling from the far end accelerates as it  propagates towards the front end and leads to final failure. 

\begin{figure}[ht]
\centering
\includegraphics[width=0.99\linewidth]{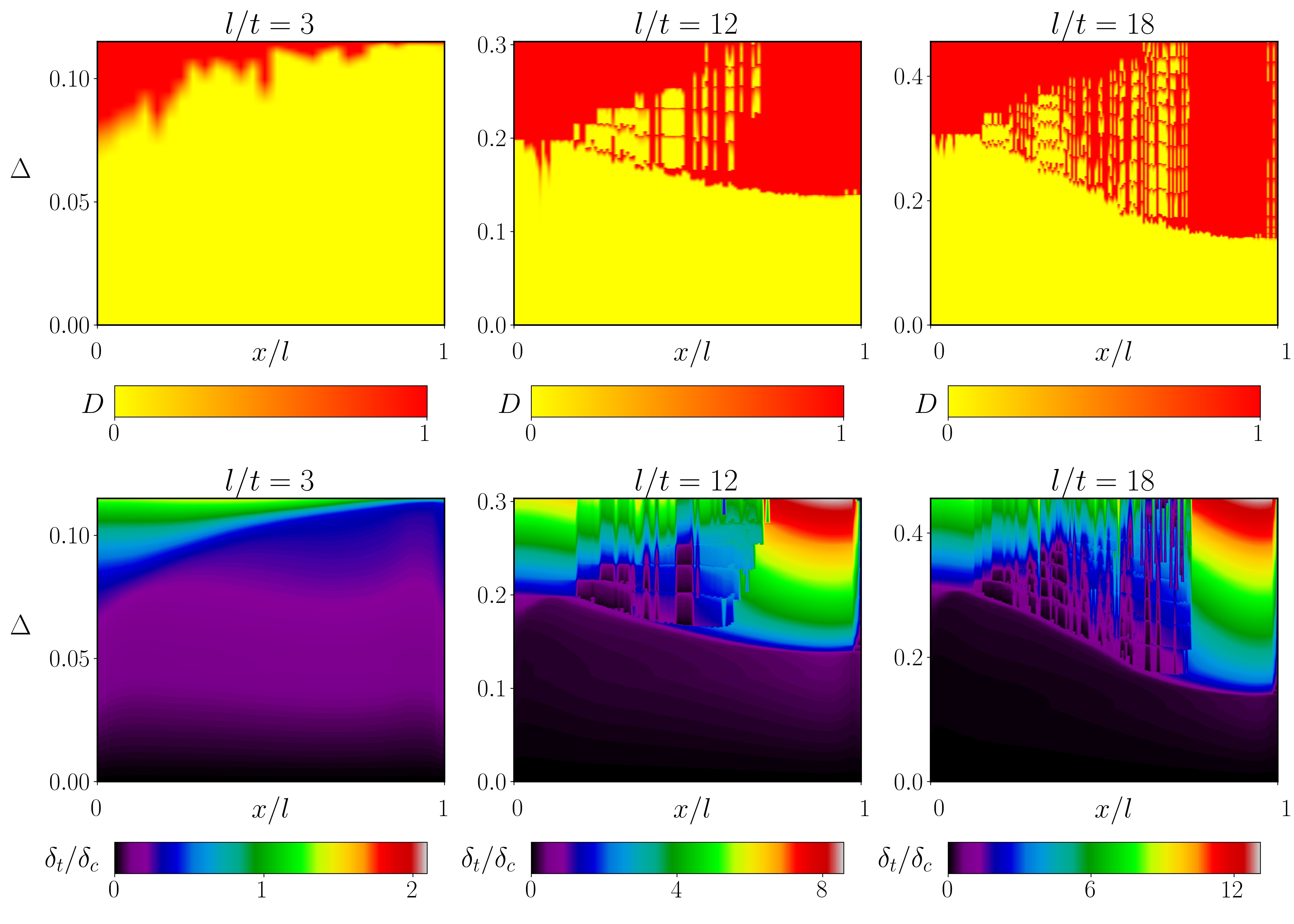}
\caption{Maps of interface damage (top row) and tangential displacement (bottom row)  for parameter values in \eqref{base_set} and for $l/t=3,12,18$. On the y-axis the imposed displacement $\Delta=u/t$ is increased, and on the x-axis the pad length $x/l$ is displayed. The damage and tangential displacements are clearly correlated and show a different picture than the normal openings.  For the larger $l/t$, formation of the interfacial cavity is sudden (occurs at a nearly constant $\Delta$). It is followed by progressive  detachment and  stick-slip events up to complete failure. In contrast, curling in the layer with $l/t=3$ occurs smoothly with no reattachment.  }
\label{fig:qualitr}
\end{figure}

An important parameter that is not visible from neither the observation (Figure \ref{exp}) nor the results in Figure \ref{fig:qualit}, is the local tangential displacement $\delta_t$. Nonetheless, the tangential motion is expected to play a significant role in determining both   the  initiation of debonding, and  the  possible occurrence of stick-slip events.  From the numerical simulations, we can quantify the tangential displacements, and the damage via \eqref{eq:damage}, as shown in Figure \ref{fig:qualitr}. From these curves the differences between curling and interfacial cavitation become even more apparent.  
An interfacial cavity of finite length forms nearly instantaneously (at a constant $\Delta$) for $l/t=12$ and $18$. This is succeeded by a more gradual propagation of damage, during which reversal of damage over time and stick-slip can be clearly observed; the sudden decrease of tangential opening indicates stick-slip as elements have reattached at a different location. In contrast, curling $(l/t=3)$ is shown to propagate gradually and smoothly without re-bonding. 
{Quite notably, the unstable propagation of  delamination in the event of curling exhibits formation of fingering-like patterns, reminiscent of the Saffman-Taylor instability \citep{saffman1958penetration}, but in the $(\Delta,x)$ plane. This morphological instability of the delamination front is highly sensitive to initial imperfections and emerges naturally in our simulations. Nonetheless, the general features of the response and the typical length-scales are consistent and independent of the mesh refinement.  }

\begin{figure}[ht]
\centering
\includegraphics[width=0.8\linewidth]{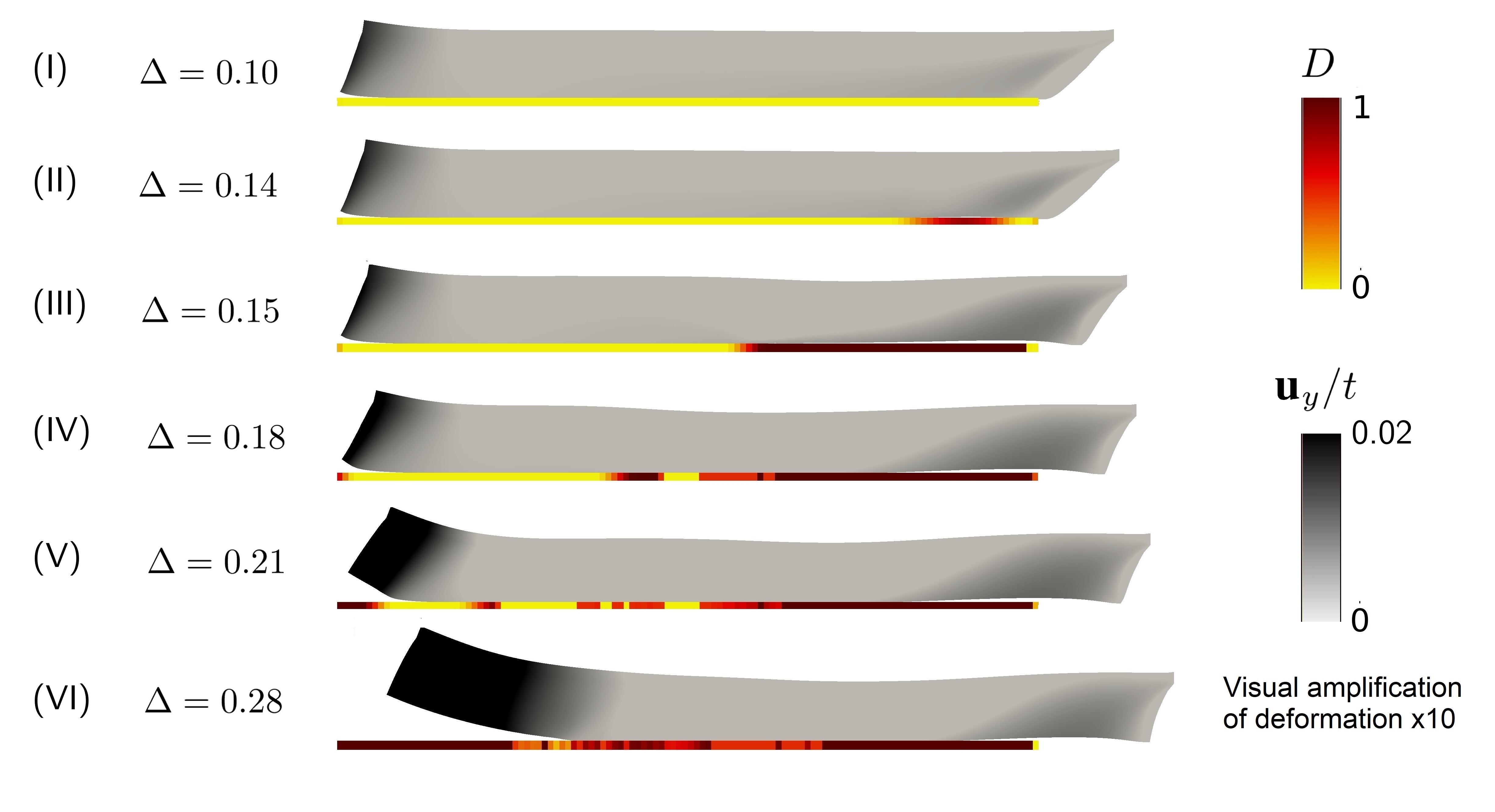}
\caption{Peeling sequence for layer with $l/t=12$  and  parameter values in \eqref{base_set}. The imposed displacement is increased from  (I) to (VI). At every stage, the (magnified) deformation of the layer is shown. Grey  shading represents vertical displacements. The shaded line at the bottom of the layer represents  the level of damage of the cohesive elements in their undeformed location. Formation of an interfacial cavity (II), and subsequent shedding of a smaller cavity (IV) are shown as reattachment occurs. Finally, curling becomes noticeable and ultimately leads to complete failure (VI). }
\label{fig:peelingsequence}
\end{figure}

To better portray the complex delamination process, which involves formation of an interfacial cavity,  we show in Figure \ref{fig:peelingsequence} a sequence of cross-sectional views of the deformed layer obtained from the plane strain simulation with $l/t=12$ (note that deformation is magnified by $10$). Here the interface (at its undeformed location) is shown as a thin line below the layer and is shaded to indicate the corresponding degree of damage. For this layer, initiation of peeling is observed first near the pulling end (II), then an interfacial  facial cavity rapidly forms (III). Shedding of a smaller cavity can be seen in (IV) by the appearance of an intermediate region that has re-adhered (yellow region in the cohesive zone). Such shedding events appear multiple times throughout the peeling process.  Finally, in (V) curling initiates; it then propagates (VI) until the layer is fully detached.

\newpage
By now, we have portrayed the peeling process in multiple ways. The normal opening, $\delta_n$, is shown to be the most intuitive field parameter and corresponds directly to our observations, however it only provides a partial understanding of the phenomena. Our numerical simulations also allow examining the tangential displacements, $\delta_t$, from which we can observe the occurrence of stick slip events, and  the damage, which provides us with clarity on the extent of failure propagation. Next, we examine how these peeling phenomena translate into load bearing capacity of the layer and determine the stability of the peeling process. 

\newpage
\subsection{Force-displacement response}

\begin{figure}[ht]
\centering
\includegraphics[width=0.99\linewidth]{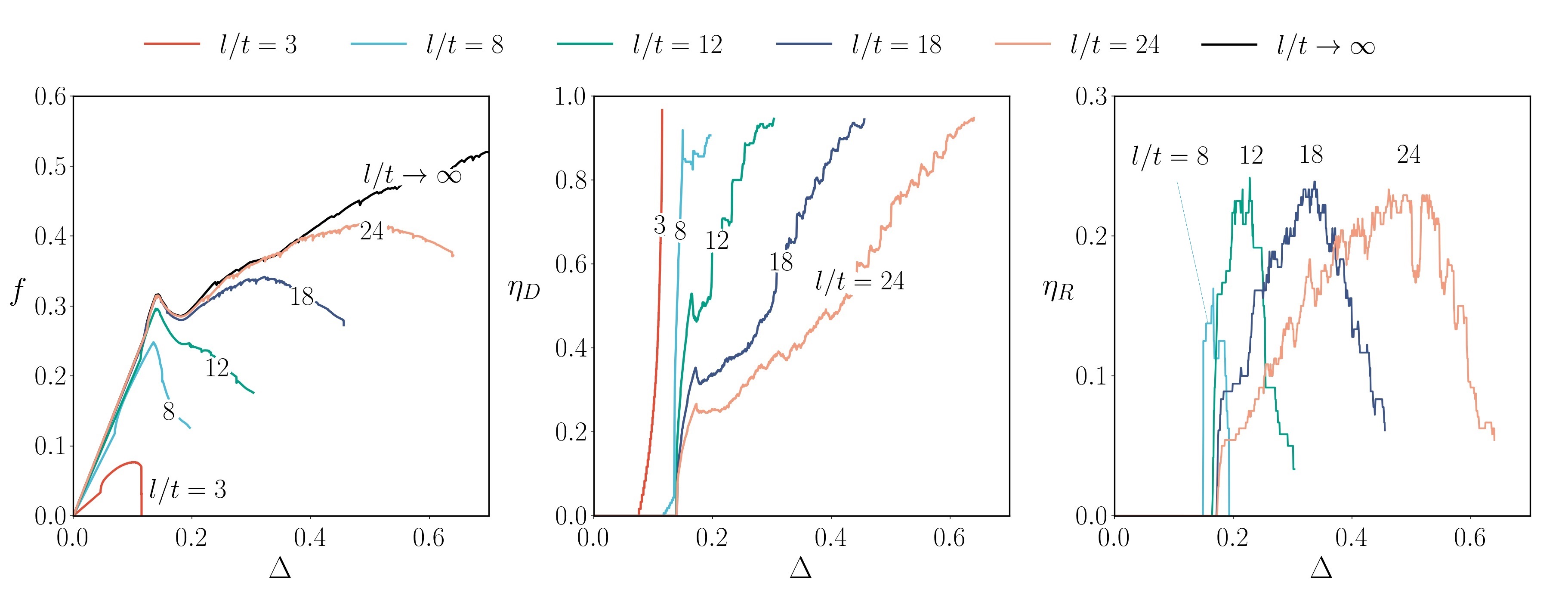}
\caption{Applied force (left), fraction of the fully broken ($D=1$) interface (middle), and fraction of reattached interface (right), shown as a function of the applied displacement $\Delta=u/t$. All curves are obtained using model parameters \eqref{base_set}, and for different aspect ratios $(l/t)$. The distinctive response of the layer with $l/t=3$ in comparison with that of larger aspect ratios is clearly observed and is indicative of the transition between curling to interfacial cavitation in conjunction with stick-slip, for increasing aspect ratios. For the infinitely long layer, $\eta_D$ and $\eta_R$ are not well defined; this layer will be discussed in detail in Section \ref{sect:infini}. }
\label{fig:force}
\end{figure}

 In Figure \ref{fig:force}  we show the variation of the  applied force, $f$, as the pulling progresses for  different aspect ratios,  and the corresponding fractions of fully detached and reattached surface areas $\eta_D$ and  $\eta_R$, respectively. 
 First, the distinct behaviour of the layer with $l/t=3$ is apparent. Failure is catastrophic as the force drops rapidly following a peak value, beyond which debonding propagates, as seen from the steep increase in $\eta_D$, and no reattachment occurs; this behavior is typical of the curling response. In contrast, for all the higher aspect ratios, the  initial slopes of the force-displacement curves is similar 
while the interface failure develops more gradually; this behavior is typical of interfacial cavitation, and the common slope can be understood by the fact that only a small region near the pulling end is activated initially. This also explains the distinct slope observed for $l/t=3$, in which case, the entire length of the pad is activated.  Interestingly, this distinction in the force-displacement response would allow to experimentally differentiate the different failure mechanisms. For larger aspect ratios, successive shedding and stick-slip events occur, as seen from the jumps in detached and reattached area.
Corresponding jumps in the force-displacement response are typical of stick-slip, and are often reported experimentally
\citep{ oldinterfaceslip, selfhealingshearpulse,  stickslipzerodegree, slidingimportant,  puresheartheory, cohen_main, debondingslip, newstudypeelinginstability}.
Reattachement persists as peeling progresses,  as seen by the increasing values of $\eta_R$, up to a peak value beyond which the entire layer is activated and a curling front begins to propagate. 


\begin{figure}[H]
\centering
\includegraphics[width=0.7\linewidth]{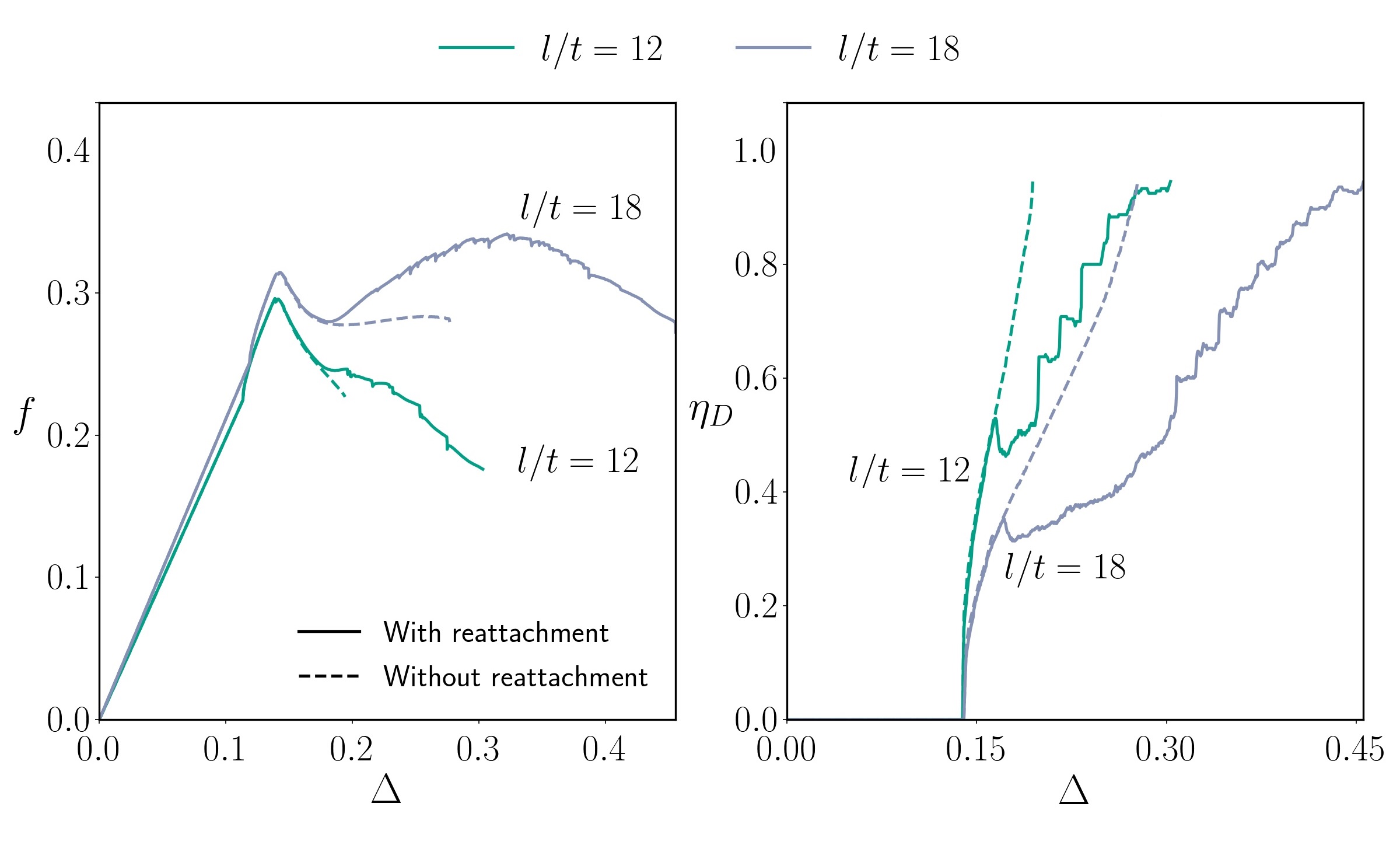}
\caption{Applied dimensionless force (left), and fraction of the fully broken ($D=1$) interface length (middle) shown as a function of the applied displacement $\Delta=u/t$. All curves are obtained using model parameters \eqref{base_set}, and for different aspect ratios $(l/t)$. The dashed force-displacement curves are for peeling without reversible adhesion.   }
\label{fig:forcenoreattachment}
\end{figure}

The response of pads with higher aspect ratios illustrates the importance of  reattachment.  Successive shedding and stick-slip events significantly impede the failure process and  increase the force-bearing capacity. Even upon initiation of  curling, rather than catastrophic failure, a slower decline in force occurs. The impact of reattachment on the response can be further explained by examining the response of  the same pads but without permitting reattachment, as shown on Figure \ref{fig:forcenoreattachment}, for layers with $l/t=12$ and $18$. 
Without reattachment, the total failure energy of the layer is reduced and failure occurs at a lower displacement and with lower applied force. Upon local initiation,  failure continues to propagate smoothly. In contrast, if reattachment can occur, it hinders the propagation and requires more energy input to induce complete failure. From this comparison, it is clear that  reversibility can have a critical influence on the overall behavior and load bearing capacity. This effect becomes more pronounced for increasingly slender layers, but appears to be bounded, by the response of an infinitely long layer, as shown by the convergence of the curves  in Figure \ref{fig:force} to that of $l/t\to \infty$. The response at this limit is further investigated in the next section. 

\subsection{Infinitely long layers and Schallamach waves } \label{sect:infini}

\begin{figure}[htp]
\centering
\includegraphics[width=0.99\linewidth]{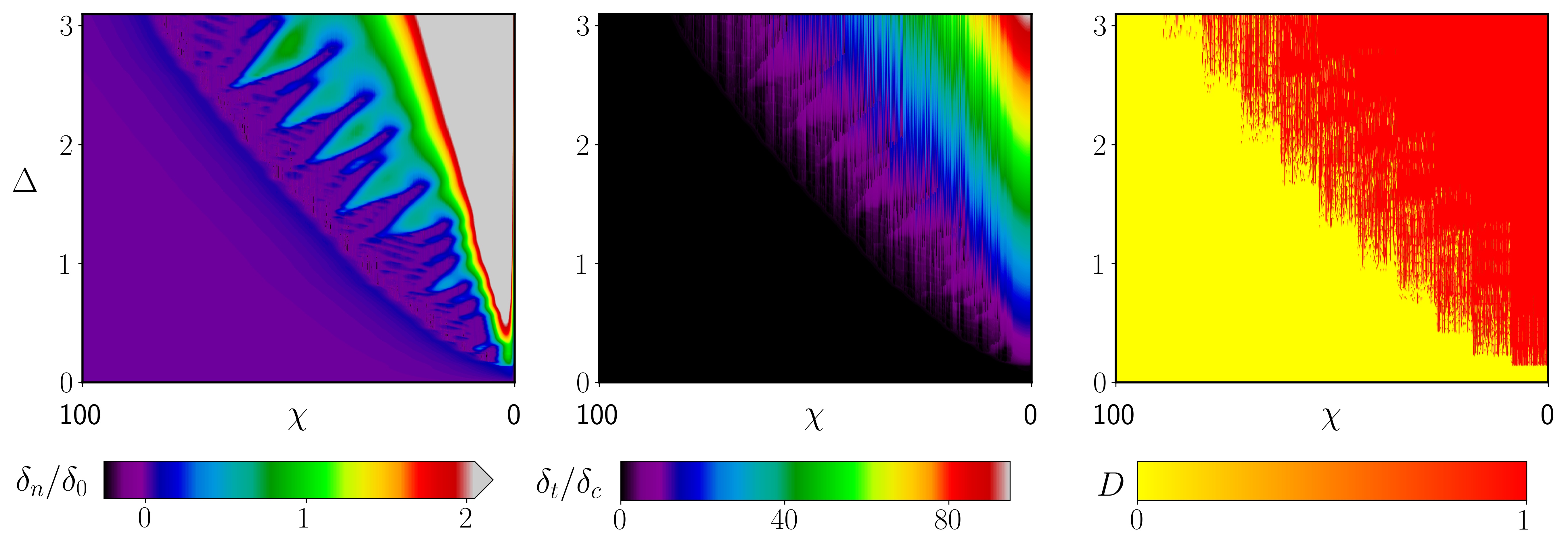}
\caption{Maps of normal opening (left), tangential displacement (middle), and interface damage (right) 
obtained using the interface values in \eqref{base_set}. On the y-axis the imposed displacement $\Delta=u/t$ is increased, and on the x-axis the pad length $\chi=(L-x)/t$ is displayed. To represent the response of an infinitely long layer $(l/t\to\infty)$, an aspect ratio of $l/t=250$ is used, and it is confirmed that no effects from the far boundary are present for the considered range of $\Delta$.  Development of a periodic response is observed after appearance of the first interfacial cavity.
}
\label{fig:infinisequence}
\end{figure}

Beyond the transitional peeling response and the formation of the first interfacial cavity, emerges a steady  peeling front that can propagate in infinitely long layers. This front is characterized by periodic events of peeling and re-adhering which become apparent by examining the map of normal opening in Figure  \ref{fig:infinisequence}.  In the wake of an ever growing interfacial cavity (grey region) a zone of periodicity develops. Within this zone, tangential displacement as well as re-adhering is present, as observed from the  corresponding maps of tangential displacement and damage. As does the interfacial cavity, this zone of periodicity continues to expand and appears to approach a steady rate of expansion (with respect to the quasi-static pulling rate), which is indicated by the linear dependence on $\Delta$.


Intuitively, the observed periodicity can be  understood by considering the displacement of the pad,  which is similar to what was shown for the layer with $l/t=12$ in Figure \ref{fig:peelingsequence}. Initially only a small region of the pad, near the pulling end, is influenced by the pulling displacement. The interfacial cavity suddenly forms  within that region and leads to the redistribution of stress that expands the  range of the affected zone. Continued pulling leads to  reattachment of an intermediate region within the cavity as it grows, and thus effectively to the shedding of a smaller cavity. Now, this reattached area locally anchors the layer to the substrate and  functions as a new pulling end. From here, the same sequence of events of back and forward motion of the peeling by formation of an interfacial cavity and reattachment, repeats periodically. The propagation of this periodic wave through the layer is captured here at the quasi-static limit. Nonetheless, its features are consistent with those of a Schallamach wave \citep{SCHALLAMACH}. {Note that numerically $l/t=250$ was used to approximate an infinite pad. It was confirmed that no end effects are present within the considered range of $\Delta$ values. }

\begin{figure}[htp]
\centering
\includegraphics[width=0.8\linewidth]{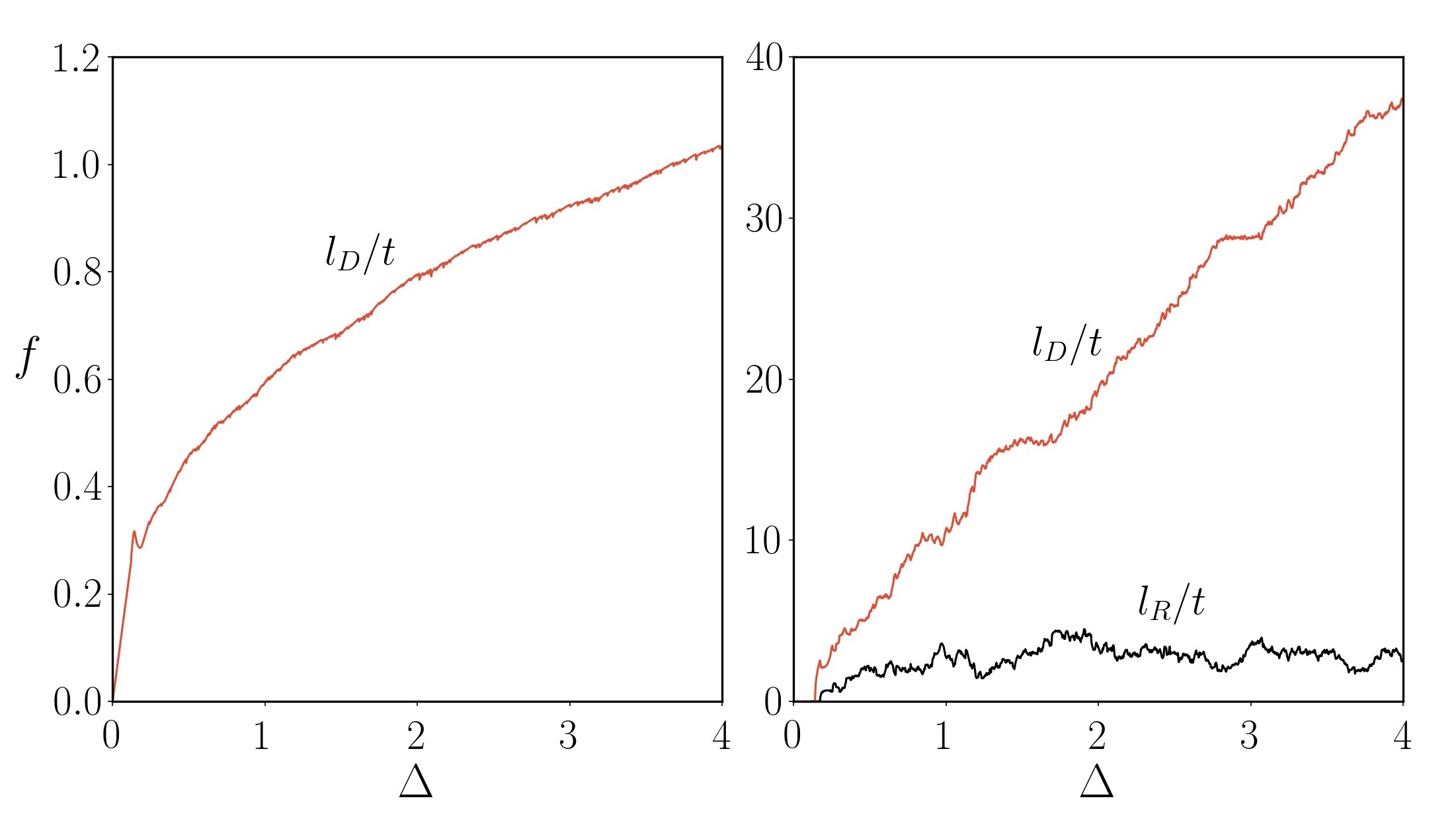}
\caption{{Applied force (left), broken interface length $l_D/t$ (elements where $D=1$) and reattached interface length $l_R/t$ (right) shown as a function of the applied displacement $\Delta$, for model parameters \eqref{base_set}. To represent the response of an infinitely long layer $l/t\to\infty$, an aspect ratio of $l/t=250$ is used. The drops and re-increases in the pulling force correspond to periodic peeling and shedding events; they are relatively less visible on this figure than on Figure \ref{fig:force} due to the larger scale of the present figure but can nevertheless be distinguished upon closer examination. }}
\label{fig:forceinfini}
\end{figure}

 The influence of the periodicity on the force-displacement response is shown in Figure \ref{fig:forceinfini} for a larger range of $\Delta$ values, along with the corresponding detached and reattached length of the interface, respectively $l_D$ and $l_R$\footnote{We display $l_D$ and $l_R$ for the infinitely long layer as $\eta_D$ and $\eta_R$ are not well defined on a infinite layer.}. Following nucleation of the first interfacial cavity (which appears as a noticeable  peak force in the early stage of pulling), drops and re-increases in the pulling force as well as the detached length correlate with the periodic peeling events. 
{The reattached length is found to saturate in this range.}

\subsection{Influence of surface properties on stick-slip} \label{sect:stickslipphase}

The stick-slip phenomena observed in the previous sections, are critically dependant on surface properties. Nonetheless, different physical  systems can exhibit a range of  surface properties that are not always straightforward to measure\footnote{For example, reported experimental values of surface energy of PDMS vary up to 3 orders of magnitude \citep{slidingimportant, strengthvalue1, strengthvalue2}.}. 

To examine this effect we consider two layers of the same  aspect ratio of $l/t=15$ and with the same dimensionless set of parameters\footnote{This set (different from earlier simulations) is chosen to allow for computational convergence within a wide range of values, which will be needed for the phase diagram that will be presented next.} \begin{equation}\label{phase_set}
\alpha=0.2,\qquad \beta=1.5,\qquad \gamma=0.96,
\end{equation}  and vary only the dimensionless surface energy $\Gamma={G_c}/{(Et)}$ and the dimensionless load-bearing capacity $\zeta={\sigma_c}/{E}$. 

\begin{figure}[ht]
\centering
\includegraphics[width=0.99\linewidth]{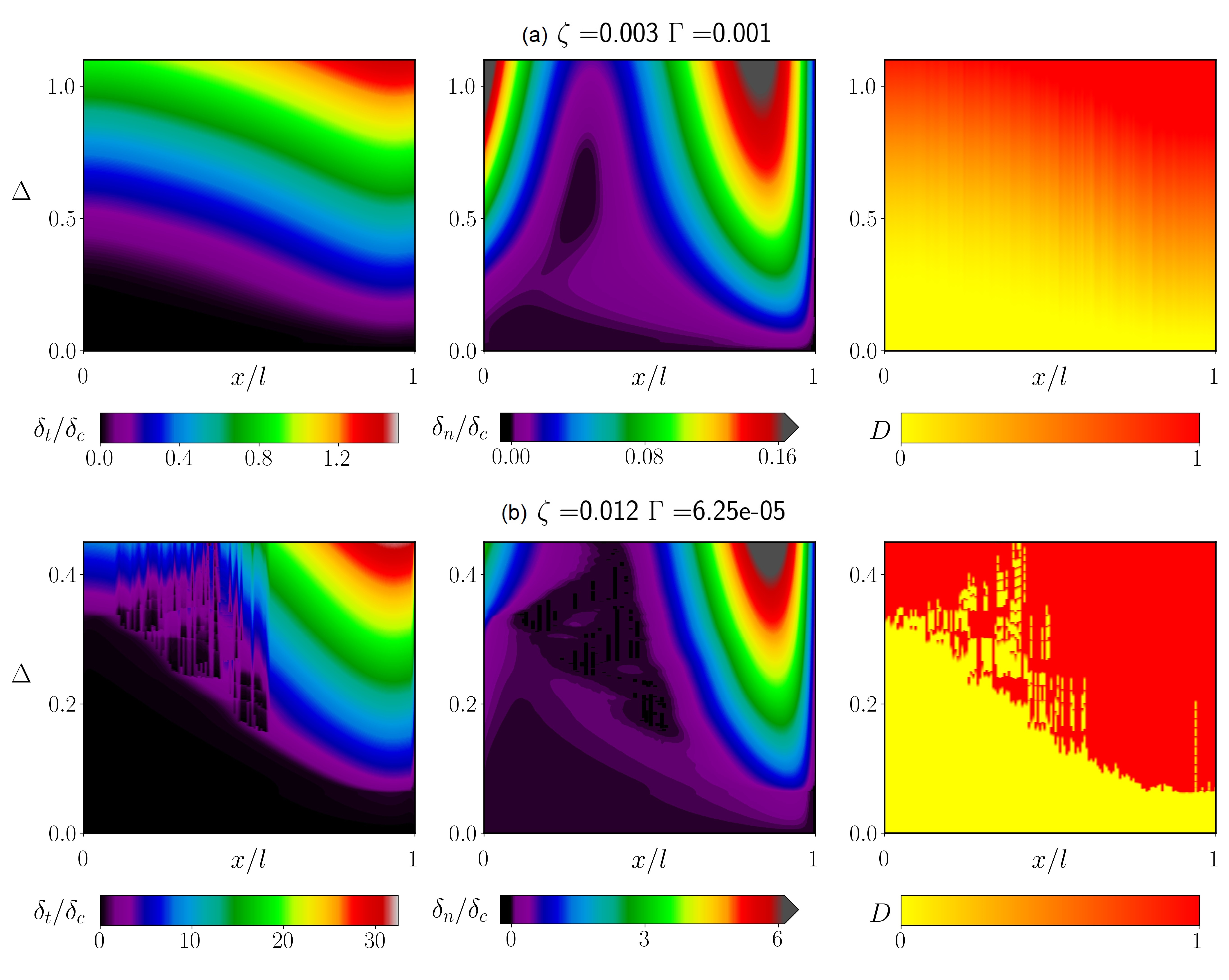}
\caption{{Maps of normal opening (left), tangential displacement (middle), and interface damage (right) 
obtained using the interface values in \eqref{phase_set}, for $l/t=15$ and for different surface parameters. On the y-axis the imposed displacement $\Delta=u/t$ is increased, and on the x-axis the pad length $x/l$ is displayed. Smooth failure along with lower displacements are observed in Figure (a), while Figure (b) shows localized highly variant displacements and damage.}
}
\label{fig:phaseopen}
\end{figure}

The surface parameters of the two layers are chosen such that one of them (a) has a very long unstable softening phase, and the second (b)  has a very short unstable softening phase.
Their vertical opening, tangential displacement and damage maps are shown in Figure \ref{fig:phaseopen}. The corresponding  force-displacement curves are included in \ref{app:forcephaseapp}. First we note that despite the apparent  similarities in the vertical displacements, layer (a) exhibits smaller displacements for similar values of $\Delta$. Next, we find that the differences appear most strikingly on the damage map. The distribution of energy within the cohesive bond in (a) allows to  distribute the stress along a longer portion of the pad before debonding. Hence a smooth transition is observed with no stick-slip events. In contrast, for (b), instability is abrupt. This leads to a higher localization of the process zone, both in space and time, and involves significant stick-slip events. The latter also shows reattachment that is not observed in case (a). The localization of displacement and damage is thus immediately linked to stick-slip and reattachment. 
This observation confirms earlier studies emphasizing the influence of the shape of the cohesive law, rather than only its energy within this range of parameters.

\begin{figure}[htp]
\centering
\includegraphics[width=0.55\linewidth]{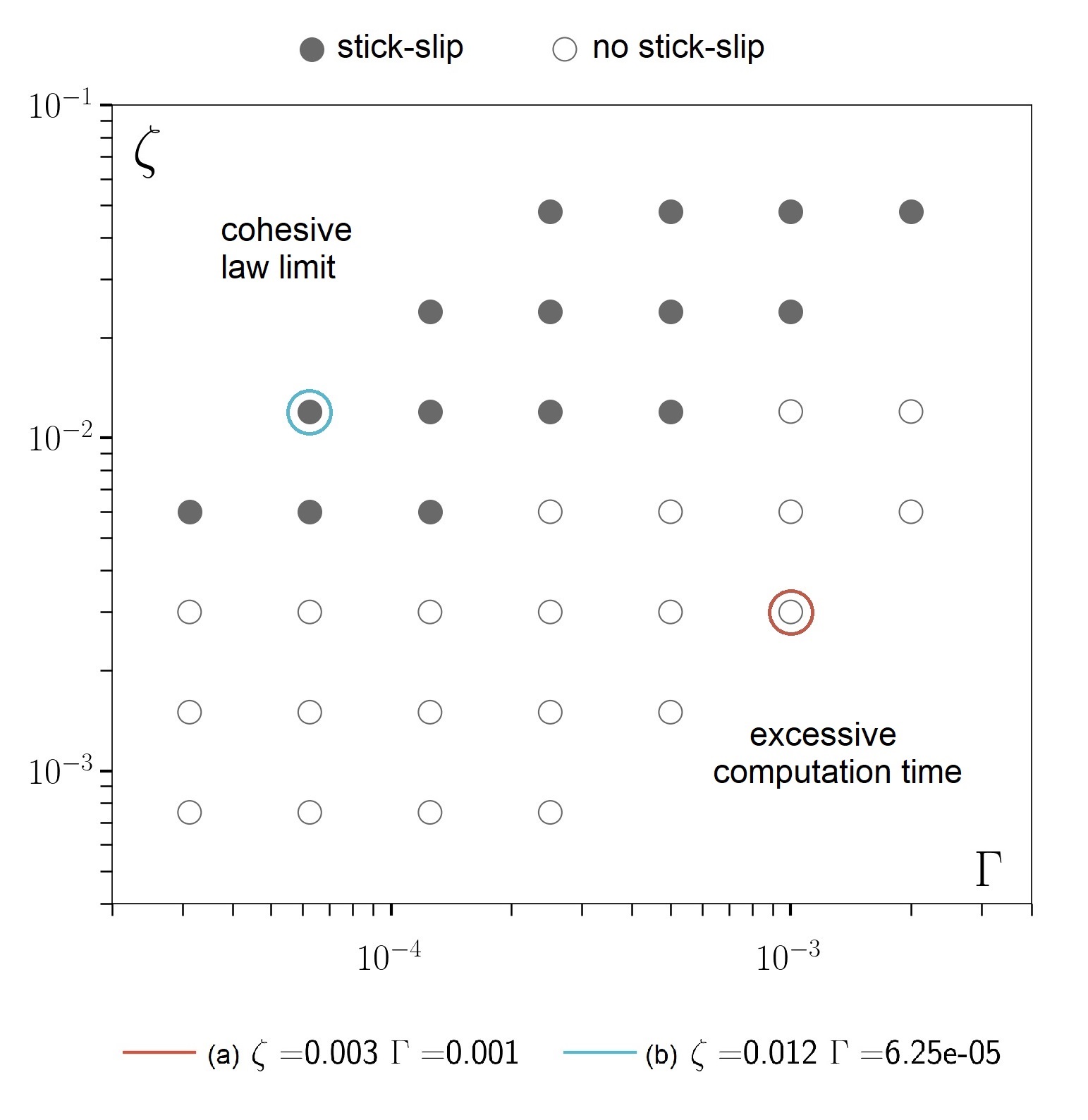}
\caption{Phase-diagram showing dependence of  stick-slip on surface properties, $\zeta$ and $\Gamma$, for layers whose remaining parameters are defined in \eqref{phase_set}. The crosses represent simulations where stick-slip occurs and the circles are for simulations where stick-slip is absent. On the top of the diagram, the parameter space is  limited by the \textit{cohesive law limit}  $\delta_c=\delta_0$. On the bottom, the  \textit{excessive computation time} refers to the parameter range for which the simulations become too computationally expensive. 
  Only curves with  consistent occurrence of stick-slip are marked.  Marked in red and blue are the two simulations that are examined in Figure \ref{fig:phaseopen}.}
\label{fig:stickslipphase}
\end{figure}

These results inspire a more comprehensive sensitivity analysis that centers on the surface properties $\Gamma$, and the specific load bearing capacity, $\zeta$, as shown by the phase diagram  in Figure \ref{fig:stickslipphase}. This diagram, which spans orders of magnitude in the parameters space, confirms that stick-slip after initial failure events is present  predominantly in layers with interface properties approaching a cohesive law with no softening phase - the cohesive law limit, at which bonds break almost instantaneously when the critical stress is achieved.

{Overall, the results in this section demonstrate the application of our modification to the classical cohesive law by \cite{ortizpandolfi}, to allow for full recovery of a cohesive bond upon contact of the two sides of the interface. It is shown that this new framework can  capture reversible interfacial peeling phenomena that emerge in various delamination processes. Specifically, this is demonstrated for elastic bi-layer adhesives capable of large deformations, pulled along the direction of the substrate.}

\section{Concluding remarks}\label{section:discussion}

From synthetic adhesive pads to tectonic plates and locomotive capabilities in nature: sliding adhesive systems exhibit various interface phenomena. In many of them, reattachment can significantly alter the observed behavior. This is especially the case for bi-layer adhesives that  peel via various failure modes, including formation of a  stable interfacial cavity near the pulling end, unstable curling emerging from the opposite end, stick-slip and shedding events, as shown  by our experimental observations. 
While earlier studies have focused on determining the conditions at initiation of failure in such systems, in this study we have developed reversible cohesive elements that are employed in a finite element framework to investigate the propagation of failure. These elements  permit re-bonding at any new location upon re-establishing contact {}{as achieved by} a straightforward adaptation to the Camacho-Ortiz cohesive law. 
{Additionally, to remedy numerical  issues that are inevitable in presence of snap-through events}, we employ a {}{specialized}  algorithm that pauses pulling during {}{spontaneous propagation of} surface failure, {thus capturing a nearly quasi-static path}.
It is shown that this simple approach is able to capture the entire range of observed behaviors. Moreover, the significant role of re-bonding is confirmed by comparing with results for which re-bonding is not permitted.  This emphasizes the importance of accounting for reversibility to study peeling processes. 

{Extending this analysis to study peeling of infinitely long layers reveals the emergence of periodic de-bonding and re-bonding events leading to the propagation of a peeling front with a steady failure sequence.} 
{This periodicity is  found to be consistent with reports of Schallamach waves albeit  in a quasi-static simulation, thus suggesting} {a rate independent} origin of Schallamach waves{, which  may be more } prevalent in a larger extent of surface interactions than first thought.
The importance of surface properties among the large range of systems where stick-slip could potentially occur, is portrayed on a phase diagram that covers several  orders of magnitude of two main surface parameters. A clear transition between systems with and without  stick-slip is shown to rely on the length of the {unstable softening phase} of the cohesive law relative to that of the stable phase. Stick-slip only occurs for interface laws with short softening phases and is linked to  localization of displacement and damage. This observation can allow to deduce the general features of a cohesive law shape based on observations of surface behavior, or inversely,  to deduce whether stick-slip will occur, based on knowledge of surface properties of a  given system. 
This work is not without limitations, future developments are needed to capture peeling processes that occur on curved surfaces. Additionally, several peeling processes can exhibit significant rate dependence, {a loss of adhesive power upon rebonding, and a dependence on previous loading and bonding states. These phenomena are not captured by the present formulation, but can be incorporated in the future, similar to \citep{wang2014crack}.}  Finally, whether it is in the study of tectonic plate {movement} or in the design of robotic devices that mimic motility in natural systems, this work emphasizes the importance of predictive models that can capture the entire peeling process as well as possible re-bonding.

\section*{Acknowledgements}
We thank Ismail Honsali and Shabnam Raayai-Ardakani for helpful conversations and for the experimental observations.  T.C. acknowledges the financial support from NSF (CMMI,
MOMS, 1942016). 
\newpage
\section*{Appendix }
\newcounter{defcounter}
\setcounter{defcounter}{0}
\newenvironment{myequation}{%
        \addtocounter{equation}{-1}
        \refstepcounter{defcounter}
        \renewcommand\theequation{A\thedefcounter}
        \begin{equation}}
{\end{equation}}

\setcounter{figure}{0}
\renewcommand{\thefigure}{A\arabic{figure}}

\setcounter{subsection}{0}
\renewcommand{\thesubsection}{Appendix A}
\subsection{Material fabrication}\label{app:fabrication}
\comment{
\begin{table}[H]
\centering
\begin{centering}
\begin{tabular}{ccc|c|c|cc}
       & \multicolumn{2}{c}{\textbf{Thickness}} & \textbf{Length} &\textbf{Depth}& \multicolumn{2}{c}{\textbf{Young's modulus}} \\
       & \textbf{adhesive}  & \textbf{backing}  &&& \textbf{adhesive}     & \textbf{backing}     \\ \cline{2-7} 
 & 6.25 $mm$          & 4 $mm$                & 31 $mm$, 88 $mm$ , 106 $mm$   & 30 $mm$ & 700 $kPa$             & 20 $MPa$    \\
\end{tabular}
\end{centering}
\end{table}}

Both the adhesive and backing were fabricated from the commercially available PDMS - Dow {\texttrademark} 184 Silicone Elastomer, using different cross-linker mass ratios to vary the elastic stiffness. For the adhesive layer  a mass ratio of $25:1$ was used, resulting in Young's modulus of $0.7$ $MPa$, and for the backing  a mass ratio of  $8:1$, resulting in a Young's modulus of $20$ $MPa$. Additional details on measurement of the material properties can be found in \citep{talcavityexpansion}. 
The adhesive mix is put in a rotating mixer for $2$ cycles of $30$ seconds, poured in an acrylic mold, de-gassed in a vacuum chamber for $3$ hours, and cured in a $40^{\circ}$ C oven for $48$ hours. This procedure is repeated for the backing, which is pored onto the adhesive layer. {The adhesive layer thus undergoes two curing cycles.} The pads are then attached to a glass plate by applying pressure manually. 
For the peeling test setup, the Instron{\texttrademark} 5943 Single Column Tabletop Testing System is used. The glass plate is clamped on one side of the machine, and the backing is clamped on the other side. The backing is pulled at a velocity of $\sim 6$ millimeter per minute.  A high-definition camera is used to capture a bottom view of the detachment of the pad through the glass plate.

\comment{
\begin{figure}[H]
    \centering
    \includegraphics[width=0.3\textwidth]{images/experimentalsetup.png}
    \caption{Illustration of the experimental procedure for execution of the pulling test. The glass to which the bi-layer pad is adhered is clamped in the 5943 Single Column Tabletop Testing Systems by Instron. The backing is clamped on the other part of the machine and is pulled. }
    \label{fig:bondstretchassumption}
\end{figure}}

\setcounter{figure}{0}
\renewcommand{\thefigure}{B\arabic{figure}}

\setcounter{subsection}{0}
\renewcommand{\thesubsection}{Appendix B}
\subsection{Numerical parameters used in the simulations} \label{app:numericalparams}

For all simulations,  thicknesses of $t=8$ $mm$   and $t_b=1.6$ $mm$ are used for the adhesive layer and the backing. The elastic modulus $E$ of the adhesive layer is $50$ $kPa$ in all simulations, except in section \ref{sect:stickslipphase} where it is $400$ $kPa$. In the numerical scheme, small levels of compressiblity are permitted by using a compressible neo-Hookean model and setting the Poisson's ratio to 0.45. The  mass density used in the simulations is $1000$ $kg/m^3$, the pulling speed is $v=0.6$ $mm/s$ and the damping time is $\theta_{1\%}$ of $0.1$ $\mu s$. Finally, the spatial discretization was found to be sufficient for a cohesive element length of $0.1t$ and the temporal discretization for a dynamic timestep of $0.1$ $\mu s$. A maximum of $28$ parallel cores are  utilized in the simulations, which 
typically takes $4-12$ hours on a high performance computing cluster.

\comment{
All simulation share the following purely numerical parameters:
\red{$nu$ and $v$ look the same  - not sure if this should be in a table}
\begin{table}[H]
\begin{centering}
\centering
\begin{tabular}{ccccccc}
 \hline \multicolumn{7}{c}{\textbf{Numerical Properties}}                                                                          \\ \hline
\textbf{$\nu$} & \textbf{$v$}        & \textbf{$\sigma_{1\%damp}$} &  \textbf{$L_{cohesive}$} & \textbf{$\rho$}&&\\
0.45           & 0.6 $mm/s$   $\times 10^{-6}$ & 0.1 $ms$                           & $t/10$ & 1000 $kg/3^3$     &&        
\end{tabular}
\end{centering}\end{table}

The following physical parameters are shared between all simulations as well:

\begin{table}[H]
\begin{centering}
\centering
\begin{tabular}{ccccccc}
\hline \multicolumn{7}{c}{\textbf{Geometrical and bulk properties}}                                                               \\ \hline
\textbf{}      & \textbf{}    & \textbf{$t$}        & \textbf{$E$}           & \textbf{$d$}         & \textbf{}               \\
               &              & 8 $mm$              & 50 $kPa$               &  1 $m$                 &                         \\
               \\           
\end{tabular}
\end{centering}\end{table}
}

\setcounter{figure}{0}
\renewcommand{\thefigure}{C\arabic{figure}}

\setcounter{subsection}{0}
\renewcommand{\thesubsection}{Appendix C}
\subsection{Comparison of force-displacement response - with and without stick-slip} \label{app:forcephaseapp}

\begin{figure}[H]
\centering
\includegraphics[width=0.75\linewidth]{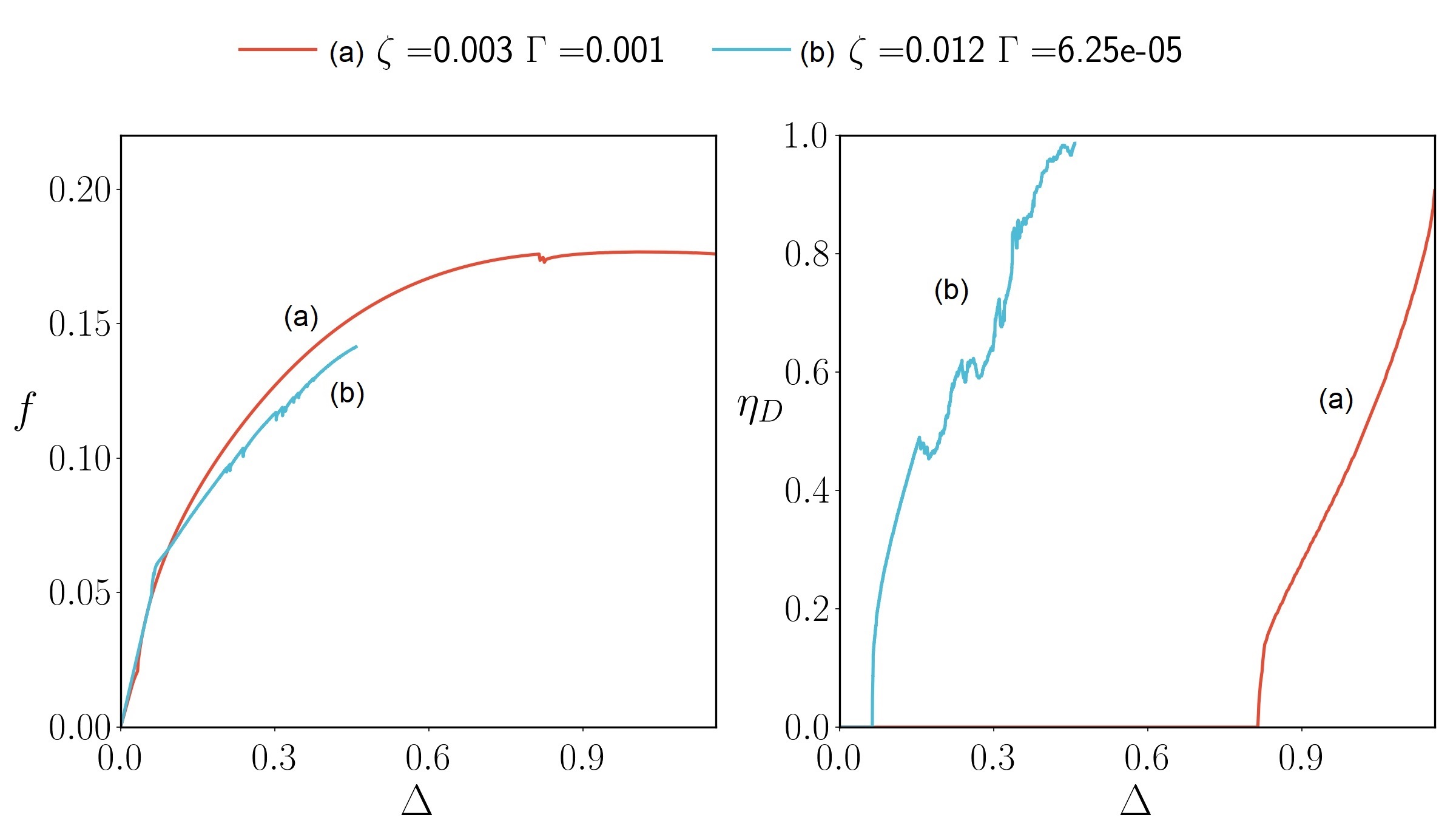}
\caption{ Applied dimensionless force (left) and fraction of broken interface (middle, elements for which $D=1$) shown as a function of the applied displacement $\Delta=u/t$, for model parameters \eqref{phase_set}, and aspect ratio $l/t=15$.}
\label{fig:phaseforces}
\end{figure}

\newpage
\bibliographystyle{elsarticle-harv}
\bibliography{ref}

\end{document}